\documentclass[aip, amsmath, amssymb]{revtex4-2}
\usepackage{natbib}
\setcitestyle{authoryear,open={(},close={)}}
\usepackage{amsmath}
\usepackage{amssymb}
\usepackage{graphicx}
\usepackage{color}
\usepackage{subcaption}
\usepackage{todonotes}
\usepackage{cleveref}
\usepackage{setspace}
\DeclareMathOperator{\proj}{\mathrm{proj}}

\begin{document}

\title{Constructing efficient score functions for rare event simulation in high-dimensional ocean-climate models}

\author{Lucas Esclapez$^{1,2}$, Val\'erian Jacques-Dumas$^{1}$, \\Reyk B\"orner$^{1,4}$, Laurent Soucasse$^{2,3}$, Henk A. Dijkstra$^{1,4}$ \\
  \small 1: Institute for Marine and Atmospheric research Utrecht, Department of Physics, \\
  \small Utrecht University, Utrecht, the Netherlands \\
  \small 2: Netherlands eScience Center, Amsterdam, the Netherlands \\
  \small 3: IMEC, Leuven, Belgium\\
\small 4: Centre for Complex Systems Studies, Department of Physics, \\
\small Utrecht University, Utrecht, the Netherlands}

\begin{abstract}

Calculating transition probabilities between different states of multistable climate tipping systems is computationally challenging in high-dimensional models.
Targeted algorithms, such as the Trajectory-Adaptive Multilevel Splitting (TAMS) method, require an adequate score function to be successful, i.e., to provide an estimate of a transition probability with an acceptable variance when only a relatively small ensemble of model trajectories can be computed.
Here, we present a data-driven method to derive a score function based on projecting the model dynamics in a reduced state space.
Using a spatially two-dimensional partial differential equation model of the Atlantic Meridional Overturning Circulation, we show that this score function performs better than currently available ones.
Using the new score function, transition probabilities can be determined with low variance, even in the case of small noise amplitudes. Besides purely noise-induced transitions, we also consider the scenario of combined stochastic and time-dependent deterministic forcing, presenting a strategy to efficiently simulate AMOC tipping events in global ocean and climate models subject to transient climate change.

\end{abstract}

\maketitle

\begin{quotation}
The Earth’s climate contains several tipping elements: subsystems that can abruptly shift from one stable state to a vastly different one. One such subsystem of primary concern is the Atlantic Meridional Overturning Circulation (AMOC), which transports heat and nutrients across both hemispheres and acts as a global conveyor belt for the global ocean circulation. While climate models suggest the AMOC could collapse due to greenhouse gas emissions and internal variability, calculating the exact probability of such a rare but catastrophic event is computationally prohibitive using a standard sampling approach. This study applies a rare-event algorithm, Trajectory-Adaptive Multilevel Splitting (TAMS), to a high-dimensional ocean model to efficiently estimate these tipping probabilities. By introducing an iterative approach to improving the score function used to bias the simulations ensemble, the study provides a reliable probability estimates for AMOC collapse under autonomous or transient forcing. This approach provides a roadmap for using rare-event algorithms to predict rare transitions in complex, state-of-the-art climate models.
\end{quotation}

\section{Introduction}

The Earth system is thought to be multistable, meaning that certain large-scale climate subsystems (so-called tipping elements~\citep{ArmstrongMcKay2022,Lenton2025}) may undergo critical transitions between distinct climate equilibrium
states.
Such tipping events could have profound regional and global impacts on nature and society.
It is therefore of critical importance for climate risk management to better understand the likelihood and underlying mechanisms of tipping events.

In principle, the internal variability of a multistable climate subsystem may be sufficient to trigger its tipping.
Mathematically, such noise-induced transitions can be modeled via stochastic dynamical systems~\citep{Hasselmann1976,freidlin_random_1998}, where the noise represents internal variability whose extremes enable transitions between different attractors of the deterministic dynamics.
In the context of climate change, however, a transition is generally triggered by the interplay between internal noise and a time-varying external forcing, which modifies the underlying stability landscape over time.
Tipping risk under anthropogenic climate change thus calls for a probabilistic assessment conditioned on a forcing protocol, such as a greenhouse gas emission scenario.

Here, we focus on the major tipping element of the global ocean, the Atlantic Meridional Overturning Circulation (AMOC).
The AMOC plays a central role in the Earth system by transporting vast amounts of heat, salt and nutrients across the Atlantic Ocean.
As demonstrated across the hierarchy of ocean-climate models, there is now ample evidence that the AMOC may lie in a multistable regime: from a conceptual model~\citep{Stommel1961} to intermediate-complexity~\citep{Rahmstorf2005} and comprehensive~\citep{vanWesten2024} Earth System Models (ESMs), a weak circulation state (AMOC-off) can coexist with the strong present-day circulation state (AMOC-on).
An AMOC collapse (transition from on to off) would have far-reaching consequences on the global climate: among others, significant cooling and modification of weather patterns over Europe~\citep{Jackson2015, vanWesten2024}, shifts in the hydrological cycle \citep{bellomo_impacts_2024}, sea-level rise over the US East coast~\citep{Little2017} and a southward shift of the Intertropical Convergence Zone~\citep{Orihuela-Pinto2022}.

Yet, the probability that global warming triggers an AMOC collapse is poorly constrained.
Until now, this problem has mostly been approached from a predictive standpoint: several recent studies~\citep{Boers2021,Ditlevsen2023,Smolders2025} have attempted to forecast the forthcoming AMOC collapse by looking for observational early warning signals.
Based on the concept of critical slowing down~\citep{Dakos2008}, these methods aim at identifying statistical indicators of an approaching tipping threshold.
While their findings suggest that the AMOC has been losing stability over the past few decades, they cannot reliably determine a tipping probability within a given time horizon.
Moreover, these methods assume that tipping occurs due to crossing a saddle-node bifurcation under slowly varying external forcing, rather than being triggered by internal variability under transient forcing.

Here, we take another view: quantifying the probability of a future AMOC collapse, conditioned on a time horizon and on a climate change scenario.
Due to the limited observational record, this task has to rely on climate modeling.
The most straightforward way of estimating a tipping probability is through direct numerical simulation (DNS).
It consists in simulating an ensemble of trajectories and counting how many ensemble members reach a collapsed state during the chosen time period.
For example, under a moderate greenhouse gas emissions scenario, \citep{Romanou2023} found a sustained AMOC collapse in two ensemble members out of ten, highlighting the sensitivity of tipping events to internal variability.
Based on their finding, one might estimate an AMOC tipping probability of $20\%$.
However, the uncertainty on this estimate is relatively high and can only be improved by using a larger ensemble, which has computational limits in state-of-the-art ESMs.
Furthermore, a sufficient sample size of tipping trajectories is necessary to obtain reliable statistics of the transition pathways and associated tipping mechanisms.
The key challenge is thus to sample AMOC transitions in high-dimensional models as efficiently as possible.

Rare-event algorithms are designed to address this challenge (see, e.g.~\citep{Cerou2019}) by biasing the ensemble simulation in a controlled way.
In the climate field, algorithms such as Giardina-Kuchan-Tailleur-Lecomte (GKTL)~\citep{Lestang2018} and Quantile Diffusion Monte-Carlo~\citep{Webber2019} have been successfully applied to generate ensembles of intense cyclones~\citep{Webber2019}, extreme heatwaves~\citep{Ragone2018, Ragone2021, Lancelin2025} or rainfall events~\citep{Wouters2023}.
GKTL has also been applied to the case of the AMOC collapse in an Earth Model of Intermediate Complexity~\citep{Cini2024}.
Among rare-event methods, Trajectory-Adaptive Multilevel Splitting~\citep{Lestang2018} (TAMS) is particularly suited to estimate probabilities.
This algorithm relies on a score function (or reaction coordinate in the field of molecular dynamics) that measures a trajectory's progress towards an AMOC collapse.
At each iteration, trajectories with the lowest scores are discarded and replaced by clones of better performing trajectories, which are then re-simulated until reaching either the time horizon or an AMOC collapse.
Once completed, TAMS yields the distribution of a large class of observables~\citep{Brehier2016,Cerou2019a}.
While powerful, TAMS is more computationally expensive than GKTL and must be optimised to be usefully applied to ESMs.

Among all parameters of TAMS, the choice of score function is the most crucial for both the efficiency and accuracy of the algorithm.
An inadequate score function results in very slow progress in TAMS iterations, even preventing convergence entirely when computational resources are limited.
Moreover, although the probability estimate is always asymptotically unbiased~\citep{Brehier2016}, its variance (i.e. uncertainty) is largely dependent on the quality of the score function~\citep{Cerou2019a}.
It is known theoretically that the optimal score function is the ``committor function''~\citep{Lestang2018,Cerou2019a}, which quantifies here the probability that a trajectory starting from a given initial condition reaches the AMOC-off state before the time horizon.

Computing the exact committor function is in general impossible, even in relatively low-dimensional systems.
There have therefore been recent efforts to approximate committor functions from simulated trajectories~\citep{Lucente2019,Finkel2021,Lucente2022,Jacques-Dumas2024}, especially in the field of computational chemistry~\citep{Lorpaiboon2024,Strahan2023,Trizio2025,Tang2024,Kang2024}.
In the case of TAMS, another difficulty arises: the committor function corresponds to the quantity we aim to compute, i.e. the transition probability.
This issue can be addressed computationally by iteratively improving the score function to get as close as possible to the committor.
To minimise data requirements (given the high computational cost), a score function improvement loop should be solely based on the trajectories sampled by TAMS .
However, until now, there are only a few attempts to couple data-driven committor estimation schemes to TAMS~\citep{Finkel2021,Lucente2022} and only one has implemented such iterative procedure in a low-dimensional system ~\citep{Jacques-Dumas2024}.

In this paper, we test the application of TAMS, combined with a score function improvement loop, on a simplified yet high-dimensional ocean model.
Our goal is to estimate, as efficiently and accurately as possible, the probability that the onset of an AMOC collapse occurs before a given time due to the combined effect of transient forcing and noise.
As a prototype model, we use a spatially resolved latitude-depth Boussinesq model of the AMOC, formulated as a stochastic partial differential equation (SPDE)~\citep{Soons2025} with around $10^4$ degrees of freedom.
This model has the advantage capturing spatio-temporal dynamics at a much lower cost compared to ESMs, facilitating the comparison of different score functions and parameter sensitivities.
Furthermore, important dynamical features, such as the most probable path of a purely noise-induced transition (the so-called instanton) are known for this model~\citep{Soons2025}.
Starting from an initial score function informed by the instanton, we design an iterative data-driven scheme based on a projection of the dynamics in a reduced state space. The scheme only uses trajectories generated by the algorithm in previous iterations.
By exploring the optimisation of TAMS under relevant constraints, we hope to inform its future application to global ocean and climate models.

We present the Boussinesq AMOC model in Section~\ref{ssec:model}, its simulation protocol in Section~\ref{ssec:protocol}
and describe the TAMS algorithm in Section~\ref{ssec:tams}.
In Section~\ref{ssec:scorefunction}, we detail our method to assess and improve the quality of the score function driving TAMS.
Then, we present our results in Section~\ref{sec:results}, first for autonomous forcing and then when applying time-dependent forcing.
Finally, we summarise and discuss our findings in Section~\ref{sec:disc}.

\section{Methodology}
\label{sec:method}

\subsection{Boussinesq model}
\label{ssec:model}
We consider an SPDE model of the meridional (zonally averaged) thermohaline circulation based on the Boussinesq approximation, forced by heat and freshwater fluxes at the sea surface.
The stability properties of the noise-free system have been extensively studied~\citep{Quon1992, Thual1992, Dijkstra1997} and it was recently studied as an SPDE by~\citep{Soons2025}.
The non-dimensionalised governing equations are recalled hereafter, but the reader is referred to~\citep{Soons2025} for a complete description of the model equations, boundary conditions and numerical implementation.

In a 2D rectangular domain, i.e., ($x$,$z$) $\in [0,A=5] \times [0,1]$ (latitude-depth), the governing equations for vorticity $\omega$, stream function $\psi$ and two tracers (salinity $S$ and temperature $T$) are:

\begin{equation}
  \begin{aligned}
    Pr^{-1} \left( \frac{\partial \omega}{\partial t} +
      \frac{\partial \psi}{\partial z}\frac{\partial \omega}{\partial x} -
      \frac{\partial \psi}{\partial x}\frac{\partial \omega}{\partial z}
    \right) &= \nabla^2\omega + Ra \left( \frac{\partial T}{\partial x} -
      \frac{\partial S}{\partial x}
    \right) \\
    \omega &= - \nabla^2 \psi \\
    \frac{\partial T}{\partial t} +
    \frac{\partial \psi}{\partial z}\frac{\partial T}{\partial x} -
    \frac{\partial \psi}{\partial x}\frac{\partial T}{\partial z}
    & = \nabla^2 T + \frac{h(z)}{\tau_T}(T_S(x) - T) \\
    \frac{\partial S}{\partial t} +
    \frac{\partial \psi}{\partial z}\frac{\partial S}{\partial x} -
    \frac{\partial \psi}{\partial x}\frac{\partial S}{\partial z}
    & = Le^{-1}\nabla^2 S + \frac{h(z)}{\tau_S}(S_S(x) + S_{S,f}(x,t) +\tilde{S}_S(x,t))\; , 
  \end{aligned}
  \label{eq:bouss}
\end{equation}

\noindent where $Pr=1.0$, $Ra=4 \times 10^{4}$, $Le=1$ are the Prandtl number, Rayleigh number and Lewis number, respectively.
The temperature and salinity surface forcings, applied via a vertical profile $h(z)$ that decays with depth, have characteristic time scales $\tau_T=0.1$ and $\tau_S=1.0$, respectively. A restoring temperature boundary condition to a symmetric (around $A/2$) $T_S(x)$ profile is applied on the surface.
The surface freshwater forcing includes an asymmetric autonomous deterministic component $S_S(x)$,
\begin{equation}
S_S(x) = 3.5\cos \left( 2\pi\left( \frac{x}{A} - \frac{1}{2} \right)\right) - \beta \sin\left( \pi \left( \frac{x}{A} - \frac{1}{2} \right)\right) \; ,
\end{equation}
where the parameter $\beta$ (fixed to $\beta=0.1$ in this paper) controls the meridional forcing asymmetry. Additionally, we consider freshwater noise $\tilde{S}_S(x,t)$,
\begin{equation}
  \label{eq:stoch_forcing}
  \tilde{S}_S(x,t) \text{d}t = \sqrt{\frac{\epsilon}{K}} \sum_{k=1}^{K}\left(\text{d}{W}_k^{(1)}(t)\cos\left(\frac{2\pi}{A}kx\right) + \text{d}{W}_k^{(2)}(t)\sin\left(\frac{2\pi}{A}kx\right)\right) \,.
\end{equation}
Here the $K$ spatial Fourier modes (we set $K=7$) are scaled by a noise strength $\epsilon > 0$ and together involve $2K$ independent Wiener processes $W_k^{(1,2)}(t)$ for $k \in \{1,...,K\}$. The noise strength $\epsilon$ is the main sensitivity parameter in this study, as it directly affects the transition probability.

We furthermore consider a time-dependent, deterministic freshwater forcing term $S_{S,f}(x,t)$ in the northern half of the domain, similar to a hosing flux added to an ESM when conducting hysteresis experiments~\citep{vanWesten2023}:
\begin{eqnarray}
  S_{S,f}(x,t) = \frac{\alpha(t)}{\sigma_{h} \sqrt{2\pi}} \exp \left(-\frac{(x-x_h)^2}{2 \sigma_{h}^2}\right),
\end{eqnarray}
\noindent where $x_h = 0.75A$, $\sigma_h = A/30$ is the hosing width parameter and $\alpha(t)$ is the time-varying hosing strength.
To ensure salinity conservation in the model, the hosing is compensated in the southern part of the domain ($x < A/2$), such that $\int_0^A S_{S,f}(x,t)\,\mathrm{d}x = 0$.

In the absence of stochastic forcing ($\epsilon = 0$) and time-dependent forcing ($\alpha(t) = 0$), the model exhibits two stable fixed points and an unstable edge state (saddle)~\citep{Soons2025}.
The stable states feature an asymmetric overturning cell located either in the Southern or Northern hemisphere.
Following~\citep{Soons2025}, we denote the state with downwelling in the northern half as the AMOC-on state (representative of the present-day AMOC), while the one with downwelling in the south is denoted as the AMOC-off state.
The edge state exhibits symmetric overturning cells, driven by the thermal boundary condition imposed on the surface.

\begin{figure*}[htbp]
  \centering
  \includegraphics[width=\linewidth]{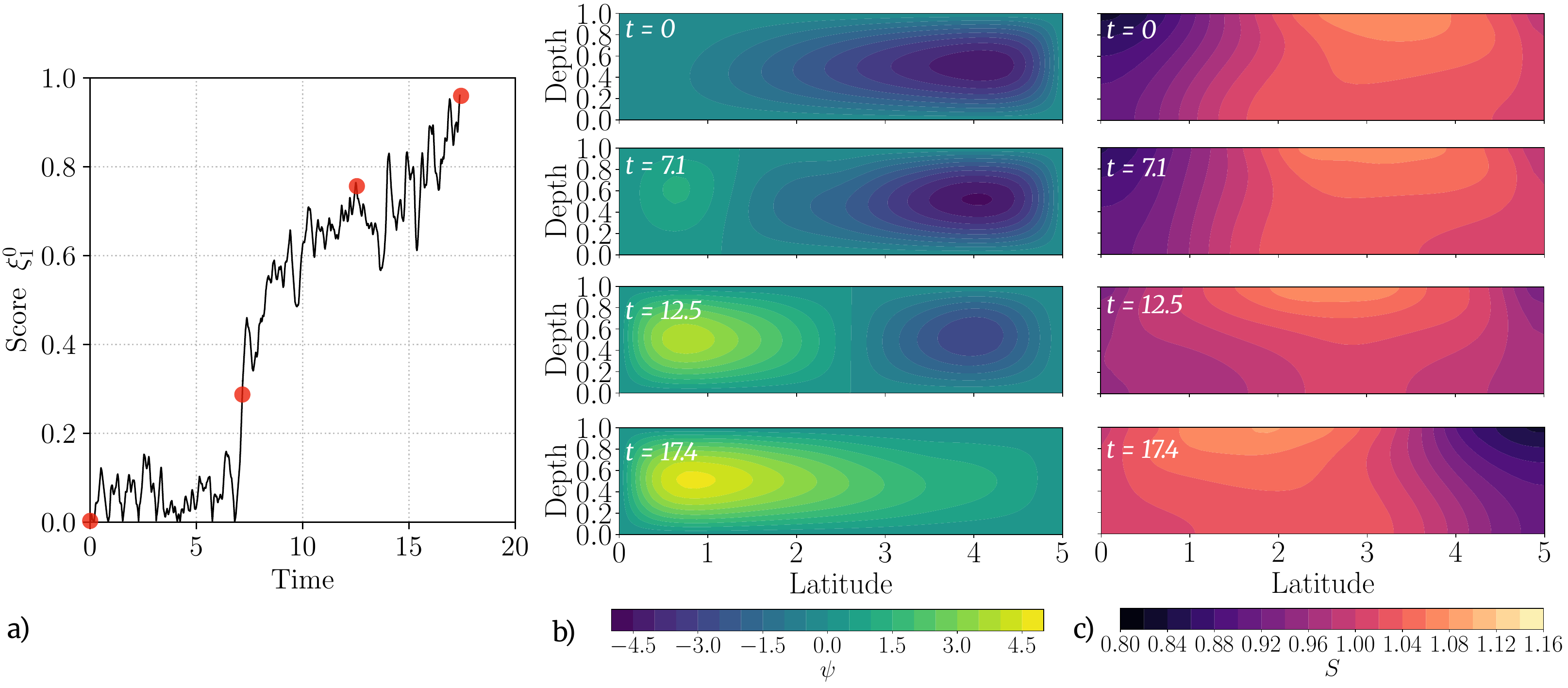}
\caption{(a) Time evolution of the score function during a noise-induced transition from the AMOC-on state to the AMOC-off state. Red dots indicate the time points of the depicted instantaneous fields; (b) stream function $\psi$ and (c) salinity field at four instants during the transition.}
\label{fig:transition_bouss}
\end{figure*}

Figure~\ref{fig:transition_bouss} shows an example of a purely noise-induced transition.
We define the AMOC strength as the stream function value measured at the center of the northern overturning cell, normalised to go from 0 in the AMOC-on state to 1 in the AMOC-off state (Fig. \ref{fig:transition_bouss}(a); this measure is further discussed in Section~\ref{sssec:ref_sf}). The time evolution of the stream function $\psi$ (Fig.~\ref{fig:transition_bouss}(b)) illustrates the reversal of the overturning cell during the transition, and the salinity field $S$ (Fig.~\ref{fig:transition_bouss}(c)) highlights the asymmetry between the AMOC-on and AMOC-off states due to the asymmetric forcing.
During the transition, the system approaches the saddle state around $t=12.5$, where two near-symmetric cells are visible with downwelling at the poles and upwelling at the equator.

\subsection{Simulation protocol}
\label{ssec:protocol}

We want to estimate the probability that the onset of an AMOC collapse occurs before a time horizon $T_a$, under the combined effects of deterministic (time-dependent, $S_{S,f}(x,t)$) and stochastic ($\tilde{S}_S(x,t)$) forcings.
From a dynamical systems point of view, the AMOC-on and AMOC-off states are both attractors.
Their basins of attraction are separated by a basin boundary on which the dynamics are locally attracted to the unstable edge state.
We restrict ourselves to the case where the deterministic dynamics stay within the bistable regime, i.e., the basins of attraction change as a function of $S_{S,f}$ but do not disappear.
The onset of the collapse can then be defined as the first entry time of a trajectory into the basin of attraction of the AMOC-off state.

Although the edge state can be computed using tracking techniques~\citep{battelino_multiple_1988, skufca_edge_2006,lucarini_edge_2017,borner_boundary_2025}, determining the location of the basin boundaries in the full phase space is typically out of reach.
However, we can determine whether a given trajectory is in the AMOC-off basin of attraction at any instant $T_a$ by continuing this trajectory while removing the stochastic forcing for $t>T_a$.
If the system is attracted towards the AMOC-off state, we can conclude that the basin boundary was indeed crossed before the time $T_a$.

This is the simulation protocol we will use here.
An ensemble of trajectories is initiated at $t=0$ on the AMOC-on state with $\alpha(0) = 0$.
From $t=0$ to $t=T_a$, stochastic forcing is applied ($\epsilon > 0$) and the time-dependent forcing $S_{S,f}(x,t)$ is linearly increased, raising the value of $\alpha$ to $\alpha(T_a) = \alpha_0$.
For a control period $t \in [T_a,2T_a]$, the simulation is then continued with a fixed value of $S_{S,f}(x,t)$ (i.e., $\alpha(t\geq T_a) = \alpha_0$) and without stochastic forcing ($\epsilon=0$).

\begin{figure}[htb]
\centering
\includegraphics[width=0.5\linewidth]{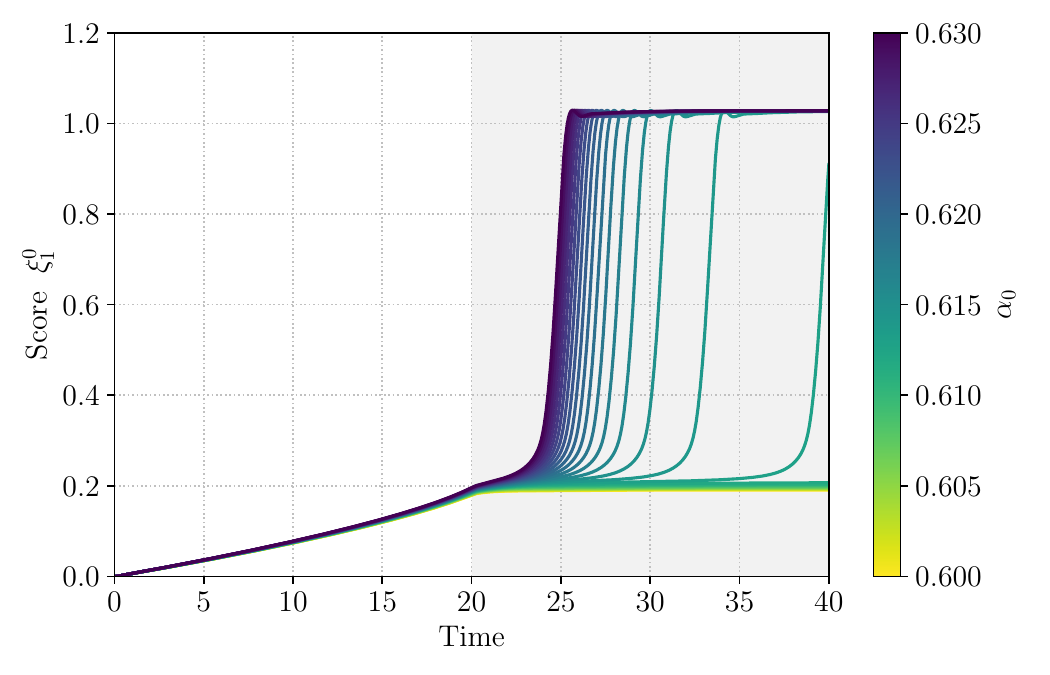}
\caption{Time series of the normalised AMOC strength $\xi_1^0$ (see Eq.~(\ref{eq:sf_naive}) below) for
trajectories under deterministic forcing with $\alpha_0 \in [0.6,0.63]$. The grey area highlights the control period.}
\label{fig:protocol_deterministic}
\end{figure}

To estimate the probability of a transition, we first study the system in the absence of stochastic forcing to determine suitable values of the hosing strength $\alpha_0$.
We set $T_a = 20$ and vary $\alpha_0$ in the interval $[0.6,0.63]$.
Past $t=23$, the trajectories for different $\alpha_0$ diverge, with a transition to the AMOC-off state ($\xi_1^0 \geq 1$) systematically occurring for $\alpha_0 > 0.612$, indicating that the basin boundary is crossed before $t = T_a$ in these cases (Fig.~\ref{fig:protocol_deterministic}).
We select values of $\alpha_0 \leq 0.6$ such that the AMOC gets close to the tipping threshold but does not tip (in the absence of noise). There is a narrow range around $\alpha_0 \approx 0.612$ for which the duration of the control period $[T_a, 2T_a]$ is insufficient to determine whether the AMOC eventually collapses or not, but the simulations reported in Section~\ref{sec:results} are all conducted for $\alpha_0$ values for which the AMOC-on state is asymptotically tracked in the absence of noise.

\subsection{Trajectory-Adaptive Multilevel Splitting}
\label{ssec:tams}
The most straightforward way of estimating the probability of an onset of the transition from AMOC-on to AMOC-off, $\mathbb{P}(\mathrm{on} \rightarrow \mathrm{off})$, is by DNS, or Monte Carlo simulation.
It consists in integrating a large number of trajectories with the protocol described in Section~\ref{ssec:protocol} and counting how many trajectories exhibit an AMOC collapse.
However, if the AMOC-on state is particularly stable, or the applied (stochastic) forcing is very weak or acts in an ineffective state space direction, or the time horizon is short, this probability may be extremely small.
Therefore, computing it by DNS may be prohibitively costly.
Moreover, even for large ensembles, DNS may only sample a few transitions, which is insufficient to compute reliable statistics of the transition time or study the drivers of the collapse.
Finally, given that the likely range of the transition probability is often unknown a priori, blindly applying DNS is risky and potentially a waste of resources.

The Trajectory-Adaptive Multilevel Splitting (TAMS) algorithm~\citep{Lestang2018} is especially fit to compute the probability that a certain event occurs before a given time horizon.
This method is a variant of Adaptive Multilevel Splitting (AMS)~\citep{Cerou2007}, which aims at estimating more generally the probability that the system enters a certain domain of its state space before another one.
To clarify the notations, consider a $d$-dimensional dynamical system described by a stochastic differential equation (SDE) of the form:
\begin{equation}
\label{eq:dynsys}
\mathrm{d}\mathbf{X}_t = f(\mathbf{X}_t)\,\mathrm{d}t + g(\mathbf{X}_t)\,\mathrm{d}\mathbf{W}_t,
\end{equation}
where $\mathbf{X}_t\in\mathbb{R}^d$ is the state vector, $f : \mathbb{R}^{d} \to \mathbb{R}^{d}$ is the drift field, and $g : \mathbb{R}^{d} \to \mathbb{R}^{d \times m}$ represents the diffusion matrix, which scales the stochastic forcing introduced by the $m$-dimensional Wiener process $\mathbf{W}_t \in \mathbb{R}^{m}$, with $m\leq d$.
Using the discretised form of the SDE, we can construct Markov chains $\mathbf{\tilde{X}} = (\mathbf{X}_t)_{t \in \mathbb{N}}$ giving the state of the system at discrete times $t \Delta_t$ ($\Delta_t > 0$ is the time step).

The Boussinesq model (Eq.~(\ref{eq:bouss})) can be recast as Eq.~(\ref{eq:dynsys}), with the discretised deterministic advection-diffusion-forcing operators assembled into $f(\mathbf{X}_t)$ and the stochastic freshwater forcing making up $g(\mathbf{X}_t)$ with a Wiener process of size $m = 2 K$.

Let $\mathcal{A}$ and $\mathcal{B}$ be two subsets of $\mathbb{R}^d$, such that $\mathcal{A} \cap \mathcal{B} = \emptyset$.
We can define the entry time into any subset $\mathcal{D}$ of $\mathbb{R}^d$ as:
\begin{align}
\tau_{\mathcal{D}} &= \mathrm{inf}\{t \in \mathbb{N} : \mathbf{X}_t \in \mathcal{D}\}
\end{align}
for a trajectory $\mathbf{\tilde{X}}$ initiated at $\mathbf{X}_{t=0} \notin \mathcal{D}$.
AMS provides an estimator $\hat{p}$ of the probability $\mathbb{P}(\tau_{\mathcal{B}} < \tau_{\mathcal{A}})$ that the system reaches the subset $\mathcal{B}$ before reaching the subset $\mathcal{A}$.
In TAMS, by setting $\mathcal{B}$ as the set of all states leading (without further forcing) to a collapsed AMOC and $\mathcal{A}$ as all states at times $t\geq T_a$, we effectively estimate the probability that the AMOC reaches the onset of collapse before the time $T_a$.

The main idea behind TAMS is to iteratively bias an ensemble of trajectories that are stopped upon reaching $T_a$ or upon entering $\mathcal{B}$.
To do that, TAMS relies on a \emph{score function} $\xi$ (further detailed in Section~\ref{ssec:scorefunction}), which measures the system's progress in approaching $\mathcal{B}$.
TAMS aims to ensure that the score of all ensemble members iteratively keeps increasing up to the global maximum $z_{\max}$, which corresponds to reaching $\mathcal{B}$.
Conversely, the score function reaches its global minimum $z_{\min}$ in the domain $\mathcal{A}$.
At each iteration, the algorithm ensures that the ensemble of trajectories is closer to $\mathcal{B}$ than at the previous iteration by discarding a certain number of trajectories, that are the furthest from $\mathcal{B}$, and replacing them by cloning and resimulating ``better performing'' trajectories (as measured by their score).
The algorithm terminates when the discarding step cannot be performed any more, i.e., when enough trajectories have reached $\mathcal{B}$ before $\mathcal{A}$ (or here, when enough trajectories have reached the onset of an AMOC collapse before $T_a$).

The main steps of the algorithm are (the complete procedure is described in~\citep{Lestang2018}):
\begin{itemize}
\item
TAMS is initialised by simulating an ensemble of $N$ unbiased trajectories $\{ \mathbf{\tilde{X}}^{(i)} \}^0_{i \in [1,N]}$, as would be the case with DNS (albeit the ensemble size $N$ is much smaller than the required number of trajectories for estimating $\mathbb{P}(\tau_{\mathcal{B}} < T_a)$ with DNS). 

\item
At each iteration $j$ of the algorithm:

\begin{enumerate}
\item
    The score $\xi(\mathbf{X}_t)$ is computed at every time step of all trajectories in the ensemble $\{ \mathbf{\tilde{X}}^{(i)} \}^j_{i \in [1,N]}$.
    For each trajectory $\mathbf{\tilde{X}}^{(i)}$, the maximum of its score is denoted as $Q^{(i)}$.
    A trajectory exhibiting a larger value of $Q^{(i)}$ than another is considered ``better performing'' because it got closer to the target subset.

\item
    Sort trajectories in order of increasing $Q^{(i)}$ and let $\mathcal{Q}^j = \mathrm{min}\{Q^{(i)}\}^j_{i \in [1,N]}$.
    The trajectories whose maximum score is $\mathcal{Q}^j$ are the ``worst-performing'' because the point where they got the closest to $\mathcal{B}$ is the smallest.
    Note that, in some cases, multiple trajectories may share the same value of $Q^{(i)}$.
    Let $l_j\geq 1$ be the number of trajectories whose maximum score is $\mathcal{Q}^j$: these trajectories are discarded.
    They are denoted with indices $d_k$, with $k\in\{1,\dots,l_j\}$.
\item
    Discarded trajectories are replaced by cloning $l_j$ trajectories selected at random in the remaining ensemble $\{ \mathbf{\tilde{X}}^{(i)}\}^j_{i \in [1,\dots, N]\setminus\{d_1,\dots,d_{l_j}\}}$.
    These trajectories are cloned up to the point where their score strictly exceeded $\mathcal{Q}^j$.
    They are then resimulated independently from that point until time $T_a$ or until reaching $\mathcal{B}$.
    This procedure ensures that $\mathcal{Q}^{j+1} > \mathcal{Q}^j$.
\end{enumerate}

\item
The algorithm is iterated until all $N$ trajectories have reached $\mathcal{B}$ or a fixed number of selection/mutation (steps 2-3 above) steps $J$ is reached.

\end{itemize}

The estimator for the probability of reaching $\mathcal{B}$ before $T_a$ is given by~\citep{Lestang2018}:
\begin{equation}
\label{eq:ams_estimator}
\hat{p} = \frac{N_\mathcal{B}}{N}\prod_{j=1}^{\tilde{J}}\left(1-\frac{l_j}{N}\right),
\end{equation}
with $N_\mathcal{B}$ the number of trajectories having reached $\mathcal{B}$ at the end of the iterative loop and $\tilde{J}$ the number of iterations performed.
At each iteration, $l_j$ trajectories out of $N$ are discarded.
Therefore, this formula can be interpreted as a product of the conditional probabilities that a given trajectory ``survives'' all iterations until reaching the target domain $\mathcal{B}$.

As shown by~\citep{Brehier2016}, the estimator $\hat{p}$ of $\mathbb{P}(\tau_{\mathcal{B}} < T_a)$ is unbiased regardless of the choice of $N$, $J$ or $\xi$.
Its variance, however, depends on the total number $N + \sum_{j=1}^{\tilde{J}} l_j$ of sampled trajectories and on the score function $\xi$.
Since $\hat p$ is a random variable, one should perform $K$ independent runs of TAMS (that can be performed in parallel) to obtain the final estimate of the rare-event probability as the mean estimator value,
\begin{align}
\overline{P}_K = \frac{1}{K}\sum_{k=1}^{K} \hat{p}_k,
\end{align}
\noindent with the associated sample variance,
\begin{align}
\sigma^2_K = \frac{1}{K-1}\sum_{k=1}^{K} (\hat{p}_k - \overline{p}_K)^2 \,.
\end{align}
The relative error can be computed as $\mathrm{RE}_K = \frac{\sigma_K}{\overline{P}_K}$.

TAMS can suffer extinction~\citep{Brehier2016, Lestang2018}, which poses a challenge because it affects the estimated rare-event probability after a finite number of TAMS runs.
The selection/mutation process is based on the deletion of all trajectories whose maximum score is one of the $k_j$ smallest values of $Q^{(i)}$.
Therefore, the number of deleted trajectories varies between iterations, and if all trajectories in the ensemble share a single value $Q^{(i)}$ (i.e., the rank of $\{Q^{(i)}\}^j_{i \in [1,N]}$ is 1), the algorithm terminates prematurely: all trajectories are discarded (extinction).
When the ensemble size $N$ is small, the repeated selection/mutation steps can quickly lead to a reduction of diversity in the ensemble until the entire ensemble is issued from a single ancestor trajectory.
If the portion of the trajectory resimulated after cloning is unable to produce a new maximum $Q^{(i)}$, extinction becomes likely.
The most effective ways to prevent this issue are to use a better score function (see Section~\ref{ssec:scorefunction}) and to increase the number of performed TAMS runs~\citep{Brehier2016}.
Extinction cannot be ruled out even when using the optimal score function, but it is proven~\citep{Brehier2016} that the estimated transition probability, averaged over $K$ TAMS runs, will eventually converge to its ground-truth value as $K$ tends to infinity.
Several techniques have been proposed to address the issue~\citep{Rolland2022, Finkel2024}, but we do not implement them here, focusing solely on the score function.
Note that when computing $\overline{P}_K$ and $\sigma^2_K$, a TAMS run that leads to extinction is counted as $\hat{p} = 0.0$.

Aside from improving the score function $\xi$ that drives the algorithm, there are two main simple ways to reduce the relative error of the probability estimate: increase the size $N$ of the simulated ensemble or increase the number $K$ of independent TAMS runs.
For a fixed computational cost, \citep{Brehier2016} advises increasing $K$ and decreasing $N$, to minimise the overall impact of extinction.

\subsection{Score function}
\label{ssec:scorefunction}

The score function $\xi$ measures the progress made in transitioning towards $\mathcal{B}$, and critically controls the selection/mutation step.
The choice of $\xi$ therefore largely determines the efficiency of the algorithm, making it the key aspect for optimising TAMS performance.

The score function affects the probability estimate $\hat{p}$ in two important ways.
First, \citep{Brehier2016} showed that $\xi$ must obey a single condition to ensure unbiasedness of $\hat p$: $\xi(\mathbf{X}_t)$ should be strictly larger than $z_{\max}$ on all points belonging to the target set $\mathcal{B}$.
Second, \citep{Cerou2019a} showed that, when $k_j=1$, $\hat{p}$ follows a central limit theorem as the size $N$ of the ensemble goes to infinity, regardless of the score function (satisfying the criteria in~\citep{Brehier2016}).
This result confirms that $\hat{p}$ is unbiased for (almost) any score function and independently of TAMS parameters.
Moreover, \citep{Cerou2019a} gives an explicit formula for the variance of the asymptotic Gaussian distribution of $\hat{p}$.
The main takeaway is that although the variance is always bounded, it strongly depends on the choice of score function $\xi$.
At worst, the probability estimate will have a variance twice as large as that of the estimate obtained with DNS: $\sigma^2 = 2p(1-p)$ (where $p=\mathbb{P}(\tau_{\mathcal{B}} < T_a)$ is the true transition probability).
At best, the variance of $\hat{p}$ is $\sigma^2=-p^2\log(p)$, which is significantly better than DNS.

The best-case variance is obtained when the score function is the committor function~\citep{Cerou2019a}, defined for any state $\mathbf{x}\in\mathbb{R}^d$ as:
\[q(\mathbf{x})=\mathbb{P}(\tau_\mathcal{B}<\tau_\mathcal{A} \mid\mathbf{X}_0=\mathbf{x}).\]

However, the committor is effectively the outcome of TAMS, and sampling this function in high-dimensional dynamical systems is computationally intractable.
In practice, the score function is therefore often a combination of the system observables, chosen based on the practitioner's intuition about the relevant physical processes, but this can be highly suboptimal and becomes increasingly difficult with growing dimensionality and complexity of the system.

Here, we introduce a data-driven score function and compare its performance against two other choices: an intuition-based score function and a score function based on the state space geometry of the equilibria~\citep{Baars2021}.
In addition, we introduce a time-dependent score function to address the specific challenge of TAMS for a simulation protocol using time-dependent forcing.

\subsubsection{Reference score functions}
\label{sssec:ref_sf}

The AMOC strength, measured as the zonally-averaged stream function value at 26$^{\circ}$N and intermediate depth, is often used to qualify the state of the AMOC in GCMs and observations~\citep{Weijer2019, Frajka-Williams2019}.
The intuition-based score function used in this work is therefore a normalised measure of the AMOC strength in the northern part of the computational domain:
\begin{align}
\label{eq:sf_naive}
\xi^0_{1}(\mathbf{X}_t) = \frac{\vert \vert \left< \psi_{\mathbf{X}_t} \right>_{N} - \left< \psi_{\mathcal{A}} \right>_{N} \vert\vert_2}
{\vert\vert \left< \psi_{\mathcal{B}} \right>_{N} - \left< \psi_{\mathcal{A}} \right>_{N} \vert\vert_2},
\end{align}
where the operator $\left< \right>_{N}$ designates an average over the $(x,z)$ domain $[0.7A,0.8A] \times [0.45,0.55]$, and $\psi_{\mathcal{A}}$ and $\psi_{\mathcal{B}}$ are the stream function of the AMOC-on and AMOC-off state, respectively.

The second reference score function was introduced by~\citep{Baars2021} to analyse a similar SPDE to the Boussinesq model studied here:
\begin{align}
\label{eq:sf_baars}
\xi^0_{2}(\mathbf{X}_t) = \eta - \eta e^{-\gamma d_\mathcal{A}(\mathbf X_t)^2} + (1 - \eta) e^{-\gamma d_\mathcal{B}(\mathbf X_t)^2},
\end{align}
where $d_\mathcal{A}$ and $d_\mathcal{B}$ are normalised distances in state space (in a suitable norm) between $\mathbf{X}_t$ and the AMOC-on and AMOC-off states, respectively. Here we choose the L2-norm of the full stream function, salinity and temperature fields.
The parameter $\gamma$ is a real positive constant arbitrarily set to 8 \citep{Baars2021} while $\eta$ is the normalised distance between the edge state and the AMOC on-state~\citep{Baars2021}.

Both $\xi^0_{1}$ and $\xi^0_{2}$ require the knowledge of the AMOC-on and AMOC-off states, and $\xi_2^0$ additionally requires knowledge of the edge state.
When considering high-dimensional models (e.g. ESMs) not amenable to continuation methods, one can use quasi-equilibrium hysteresis experiments~\citep{vanWesten2023} to find the stable states. Determining the edge state requires e.g. edge tracking techniques \citep{borner_boundary_2025}.

\subsubsection{Data-driven score function \label{ssec:data-driven-score}}

\begin{figure*}[htb]
\centering
\includegraphics[width=\linewidth]{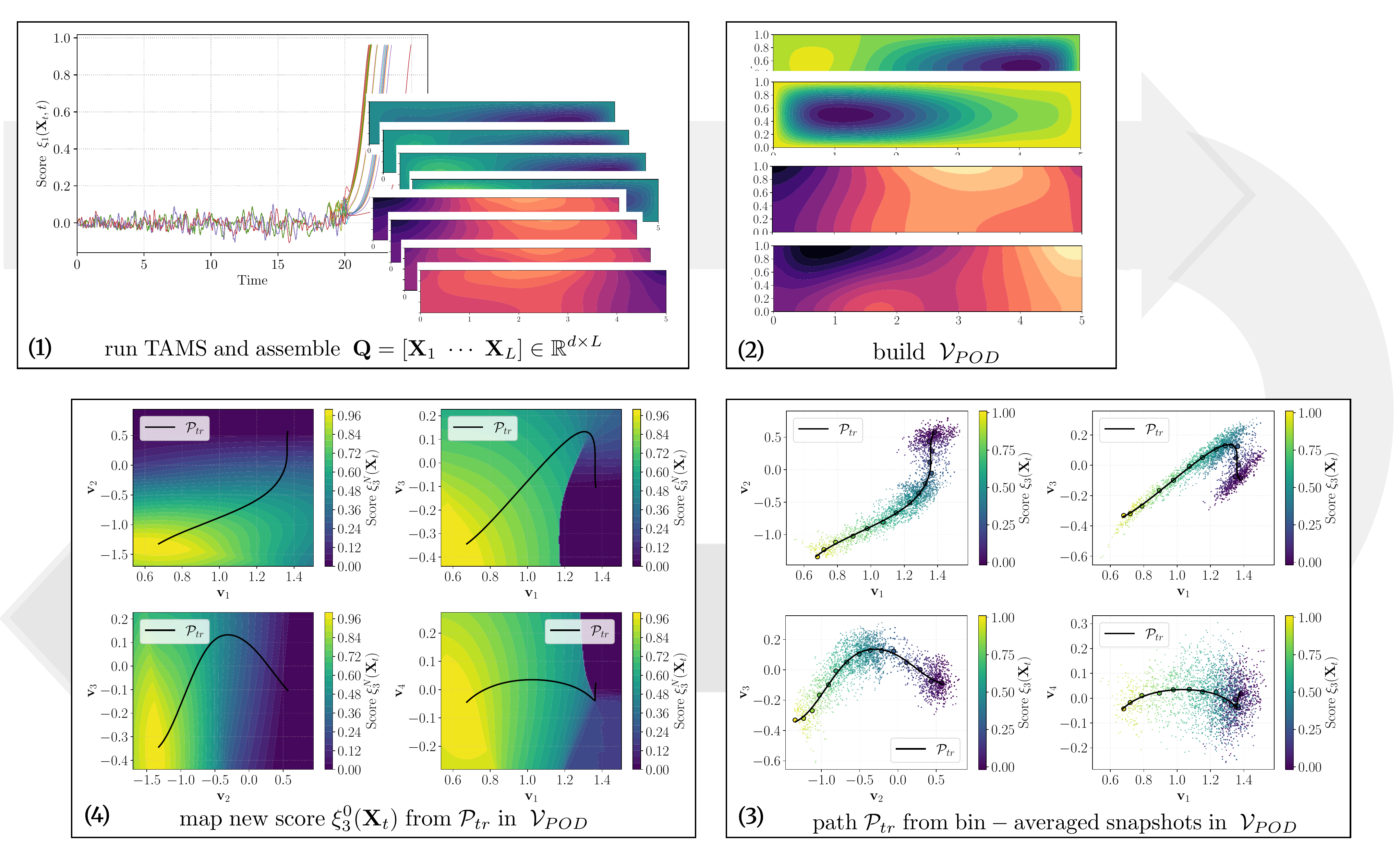}
\caption{Sequential steps in constructing the score function $\xi^0_{3}$ from a TAMS run: 1) Assemble the data matrix $\mathbf{Q}$ from the ensemble of trajectories obtained after a TAMS run, 2) perform POD to construct the latent space $\mathcal{V}_{POD}$, 3) project $\mathbf{Q}$ in $\mathcal{V}_{POD}$ and construct the ensemble transition path $\mathcal{P}_{tr}$, 4) map the full latent space with a score function $\xi^0_{3}$ using the arc-length along $\mathcal{P}_{tr}$.}
\label{fig:ScoreLoopIteration}
\end{figure*}

To construct an interpretable, data-driven score function, we perform a linear dimensionality reduction using Proper Orthogonal Decomposition (POD)~\citep{Berkooz1993}, an approach widely employed in fluid dynamics (also referred to as EOF in the climate community).
This dimension reduction technique is combined with a non-linear path reconstruction method akin to Principal Curve~\citep{Hastie1989}, to provide a score map in the POD latent space (i.e., a reduced state space).
This second step is close to what was suggested for TAMS by~\citep{Wang2021}, but it has not been combined with dimensionality reduction and here we perform binning in the 1D score function space instead of the model phase space.
The entire process is described below, from running TAMS and gathering data to mapping a score function (Fig.~\ref{fig:ScoreLoopIteration}).

To apply POD, we assemble a data matrix $\mathbf{Q} = [\mathbf{X}_1\ \cdots \ \mathbf{X}_L] \in \mathbb{R}^{d \times L}$ from $L$ independent model states $\{\mathbf{X}_i\}_1^L \in \mathbb{R}^d$.
We then compute the singular value decomposition of $\mathbf{Q}$,
\begin{equation}
\mathbf{U} \mathbf{\Sigma} \mathbf{V}^T = \mathbf{Q},
\end{equation}
where $\mathbf{U} \in \mathbb{R}^{d \times d}$ is an orthogonal matrix containing the left singular vectors, $\mathbf{\Sigma} \in \mathbb{R}^{d \times L}$ is a rectangular diagonal matrix and $\mathbf{V} \in \mathbb{R}^{L \times L}$ an orthogonal matrix containing the right singular vectors.
The diagonal elements of $\mathbf{\Sigma}$ are the singular values of $\mathbf{Q}$ arranged in decreasing order.
The POD basis matrix $\mathbf{U}_{l} \in \mathbb{R}^{d \times l}$ is then defined to be the $l$ left singular vectors associated with the leading $l$ singular values:
\begin{equation}
\mathbf{U}_{l} = \big[\, \mathbf{u}_1 \ \mathbf{u}_2 \ \cdots \ \mathbf{u}_l \,\big]
\in \mathbb{R}^{d \times l}.
\end{equation}

In the following, the $l$ left singular vectors are referred to as the POD modes.
By construction, the dynamics encoded in these POD modes contain most of the energy of $\mathbf{Q}$.
The associated $l$-dimensional embedding of $\mathbf{Q}$ is then given by the projection of the state data onto this reduced basis:
\begin{equation}
\mathcal{V}_{POD} = U_l^{T} Q \in \mathbb{R}^{l \times L},
\end{equation}
so that each $d$-dimensional snapshot $\mathbf{X}_t$ is represented by a low-dimensional coordinate vector $\mathbf{v}_t$  with components $v_i = U_l^{T}\mathbf{X}_i$.

\begin{figure*}[htbp]
\centering
\includegraphics[width=\linewidth]{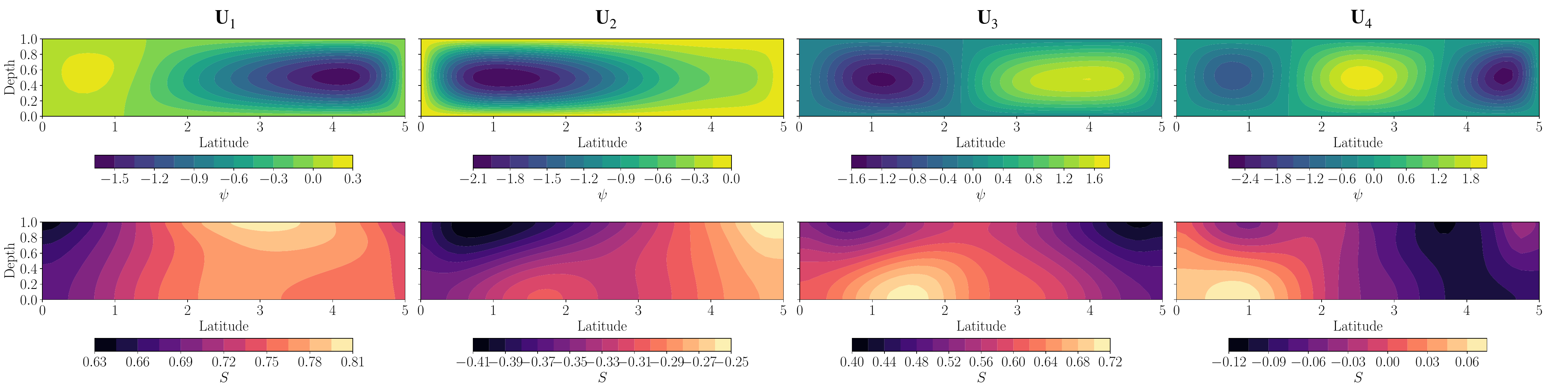}
\caption{POD modes $\mathbf{U}_{1-4}$ from left to right, for the stream function $\psi$ (top row) and salinity $S$ (bottom row).}
\label{fig:PODmodes}
\end{figure*}

Here, we apply the POD decomposition to the stream function $\psi$ and salinity field $S$ ($d=6642$; we omit the vorticity and temperature data). Both fields are scaled to ensure that they contribute equally to the total "energy" content of the system (in terms of L2 norm).
Model states are also spatially weighted such that the scalar product of two state vectors corresponds to a spatial average.
In the Boussinesq model, we found that at most $l = 8$ POD modes are needed to represent $98\%$ of the energy contained in $\mathbf{Q}$.
Three dominant modes add up either to the AMOC-on or the AMOC-off state, with two different sets of weights.
All other modes are only non-negligible during the transition.
The exact shapes of the modes and the order in which they are ranked (after the first three modes) depend on the data contained in $\mathbf{Q}$ (see Fig.~\ref{fig:PODmodes} for a typical reconstruction of the first four modes).
Higher modes are mostly multipole in the latitude direction in response to the harmonic forcing imposed on the surface (see Eq.~(\ref{eq:stoch_forcing})).

In practice, the data matrix $\mathbf{Q}$ contains model states from two sources: 1) states from all active trajectories at the end of a TAMS run (i.e. the trajectories that effectively transitioned), 2) states from two statistically stationary trajectories entirely simulated in the AMOC-on and AMOC-off basin under low noise conditions.
The latter are necessary to ensure that $\mathbf{Q}$ encompasses a sufficient representation of AMOC-on and AMOC-off states, regardless of the outcome of TAMS runs.
The total sample size $L$ varies, depending on the result of the TAMS run, with $6000 < L < 10000$ and each of the stationary trajectories containing 200 time points.
As an alternative to building $\mathbf{Q}$ from TAMS data, data from a quasi-equilibrium experiment (increasing $\alpha(t)$ very slowly until a collapse is observed) can be used.
This approach is computationally less expensive because it involves only a single, long trajectory, but we find the resulting score function to under-perform compared to using TAMS data (see Appendix~\ref{sec:data_driven_tests}).

Once the POD latent space has been constructed, we can build a data-driven score function in this low-dimensional space.
The main idea is to track in the POD space the different transition pathways sampled thus far by TAMS.
The orientation of the isolevels of the score function is derived from a weighted average of these paths, the mean transition path (MTP, denoted $\mathcal{P}_{tr}$), multiplied by an exponentially decaying kernel as the system deviates from $\mathcal{P}_{tr}$.
After each TAMS run, the score function is iteratively improved based on the new simulation data.

First, all model states of $\mathbf{Q}$ are projected onto $\mathcal{V}_{POD}$.
This way, we obtain a point cloud of the subspace explored during the TAMS run (as well as the AMOC-on and AMOC-off states).
Each state is labelled with its current score function value, such that data points in the basis $\mathcal{V}_{POD}$ can be clustered into $N_b$ bins between $z_{\min}$ and $z_{\max}$.
In each bin, we use a k-mean algorithm with $k \in \{1,2\}$ to identify either one or two clusters.
In the first case, we conclude that all data points follow the same transition pathway.
But if we find two clusters with barycenters sufficiently separated (above 1.5 times the standard deviation of the data points in the current bin), we assume that two separate pathways have been detected.
Both pathways will then be tracked separately from the next bin on.

After transition paths have been tracked across all $N_b$ bins, MTPs are constructed by fitting smooth parametric B-splines through the $N_b$ bin barycenters~\citep{Dierckx1982}.
The arclength $s$ along the paths, normalised between 0 and 1, is then used to build the score function $\xi^0_{3}$ as a distance-weighted average of the different paths.
The system state $\mathbf{X}_t$ at time $t$ is first projected in $\mathcal{V}_{POD}$ and then onto each reference path $\mathcal{P}_{tr}$. We apply an exponential decay of the score transversal to $\mathcal P_{tr}$~\citep{Wang2021}, giving
\begin{equation}
    \begin{aligned}
        \xi^0_{3}(\mathbf{X}_t) &= \frac{\sum_{tr=1}^{n_{path}} s(\mathbf{v}_t, \mathcal{P}_{tr}) w_{tr}(\mathbf{v}_t)}{\sum_{tr=1}^{n_{path}} w_{tr}(\mathbf{v}_t)}\\
        w_{tr}(\mathbf{v}_t) &= \exp\left( -\frac{d(\mathbf{v}_t, \mathcal{P}_{tr})^2}{d_0^2}\right) \,.
    \end{aligned}
    \label{eq:sf_pod}
\end{equation}
Here $\mathbf{v}_t = \proj_{\mathcal{V}_{POD}} \left(\mathbf{X}_t \right)$ and $s(\mathbf{v}_t, \mathcal{P}_{tr})$ is the value of $s$ at the closest point of $\mathcal{P}_{tr}$ from $\mathbf{v}_t$.
The distance between the closest point of $\mathcal{P}_{tr}$ and $\mathbf{v}_t$ is denoted $d(\mathbf{v}_t, \mathcal{P}_{tr})$.
We set $d_0=1$, which is of the same order of magnitude as the largest of the ranges of $\mathcal{P}_{tr}$ projected on the latent space basis.
The effect of $d_0$ on the performance of the score function is discussed in Appendix~\ref{sec:data_driven_tests}.

At the initial score function iteration, we build $\mathbf{Q}$ using the data obtained from a TAMS run driven by $\xi^0_{1}$, using a large noise level $\epsilon$ to make transitions likely to occur.
Over a few iterations, the noise level is reduced to the desired level while a refined version of $\xi^{0}_{3}$ is generated each iteration.

To produce the score $\xi^0_{3}$ employed in Section~\ref{sec:results}, we performed TAMS in an autonomous setting ($\alpha_0 = 0$), selecting an initial noise amplitude $\epsilon = 0.08$.
The noise level was geometrically reduced with a ratio $r=0.8$ for the first 9 iterations, until reaching the lower end of our range of interest. While we only found a single transition path in the Boussinesq model using $\xi_3^0$, our method works also for systems with multiple transition channels (see Appendix \ref{sec:test_datascore}).

Figure~\ref{fig:ensemblepath_inPOD} shows all the MTPs generated over the multiple iterations and projected in the POD space obtained at the last iteration.
Overall, the average transition path is only moderately affected by the noise amplitude, as demonstrated by the proximity of all paths in the two dominant two POD modes (Fig.~\ref{fig:ensemblepath_inPOD}a).
Modes of lower importance show more relative changes as iterations proceed (Fig.~\ref{fig:ensemblepath_inPOD}b, c).

It is interesting to compare the MTP, i.e. the expected transition path under the given noise, with the most likely transition path in the limit of vanishing noise, given theoretically by the Freidlin-Wentzell (FW) instanton~\citep{freidlin_random_1998}.
This instanton was computed by~\citep{Soons2025} for the Boussinesq model and the present noise structure.
As seen in Fig.~\ref{fig:ensemblepath_inPOD}, the instanton and the final MTP join near the edge state and remain close to each other in the basin of attraction of the AMOC-off state.
By contrast, the two paths differ significantly in the basin of the AMOC-on state, with the FW instanton exhibiting a circular excursion not followed by the MTP.
This difference may be interpreted as a finite-noise effect, by considering a correction that accounts for the divergence of the drift field \citep{borner_saddle_2024} (see Appendix~\ref{sec:instanton_vs_mtp}).

\begin{figure*}[htb]
\centering
\includegraphics[width=\linewidth]{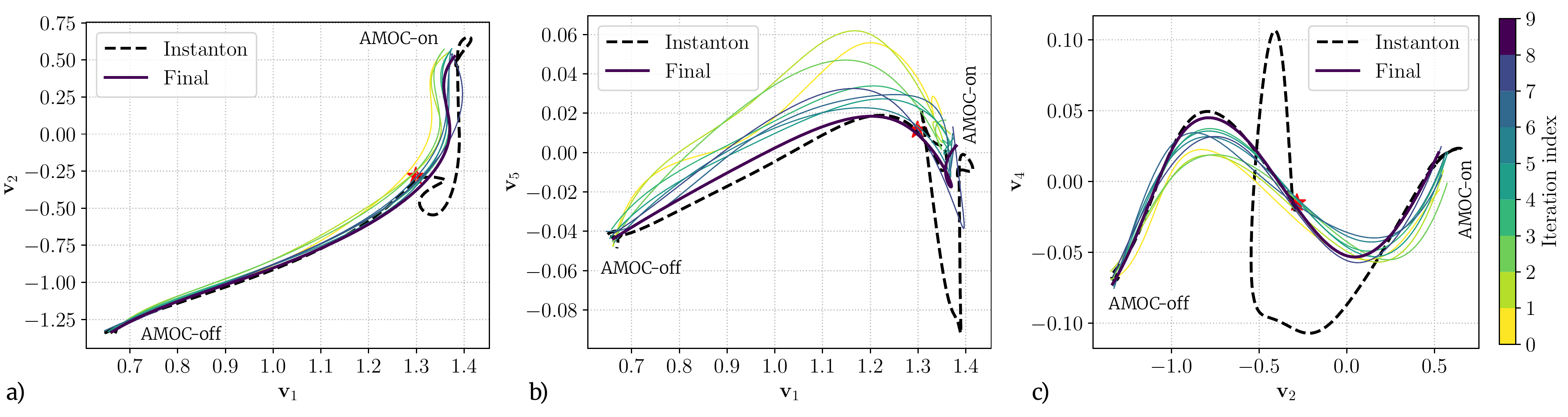}
\caption{Mean transition paths (MTPs) (full lines) obtained during the score function improvement iterations, projected in three pairs of POD modes. The FW instanton (dashed line) and the saddle state (red star) are added for reference.}
\label{fig:ensemblepath_inPOD}
\end{figure*}

\subsubsection{Time-dependent score function}

\begin{figure}[htb]
\centering
\includegraphics[width=0.5\linewidth]{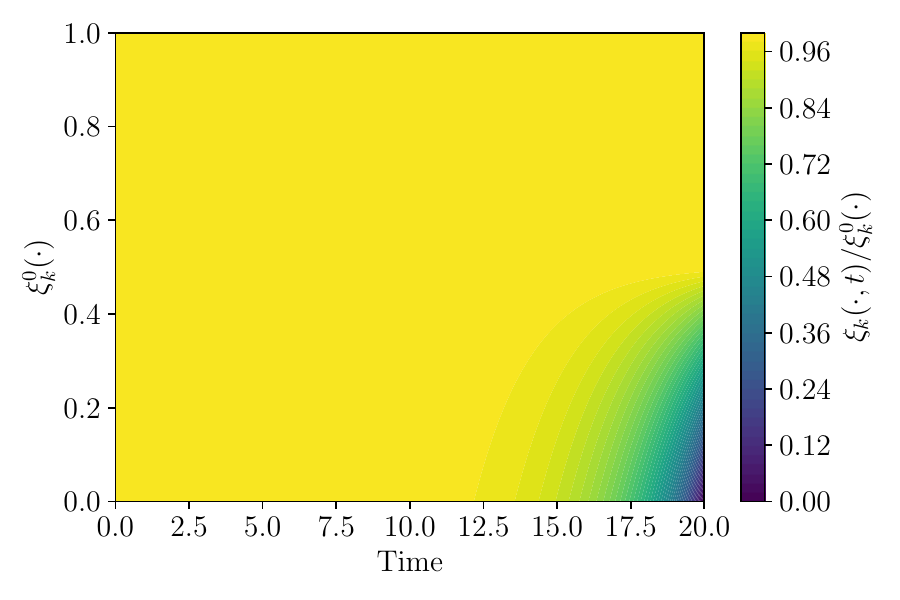}
\caption{Ratio of the dynamic $\xi_k(\cdot,t)$ to the static $\xi^0_k(\cdot)$ score function with $T_a = 20$, $\tau_s = 2.0$ and $\xi^0_k(\mathbf{X}_{edge}) = 0.5$.}
\label{fig:dynamic_scorefunction}
\end{figure}

So far, we have introduced three score functions that only depend on the system state $\mathbf{X}_t$ but not on time $t$.
They will be referred to as \emph{static}, and have a superscript $0$.
However, when using TAMS, we are computing the probability that an event occurs before a certain time horizon $T_a$.
Therefore, the associated committor function depends on both the system state and time.
\citep{Lestang2018} showed that, for a simple Ornstein-Uhlenbeck process, the time-dependent committor is well approximated by the static committor, except in a thin spatial boundary layer of size $\epsilon$ near $\mathcal{B}$ and a time boundary layer of size $\tau_s$ near the time horizon $T_a$.
The thickness $\tau_s$ is the correlation time of the system.
We introduce dynamic versions of the three static score functions (Sect.~\ref{sssec:ref_sf} and \ref{ssec:data-driven-score}) as follows:
\begin{align}
\label{eq:dynamic}
\xi_k(\mathbf{X}_t,t) = \xi^0_k(\mathbf{X}_t) \left(1.0 - e^{\frac{t - T_a}{\tau_s}} \cdot \frac{\mathrm{max}\left(0, \xi^0_k(\mathbf{X}_{edge}) - \xi^0_k(\mathbf{X}_t)\right)} {\xi^0_k(\mathbf{X}_{edge}) - \xi^0_k(\mathbf{X}_t(t=0))}\right),\ \ \ k\in\{1,2,3\}.
\end{align}
This introduces an exponential decay near the time horizon $T_a$, with a characteristic time scale $\tau_s$.
The decay is scaled by the remaining progress towards the target value of the score function at $T_a$.
In the simulation protocol (Sect.~\ref{ssec:protocol}), our objective is to reach the basin boundary before $T_a$, so our target score value is set here to $\xi^0_k(\mathbf{X}_{edge})$, the value of the score function at the edge state. For most values of $\xi_{k}^0$ and $t$, the dynamic and static score functions have a similar value (Fig.~\ref{fig:dynamic_scorefunction}).

When the dynamical system is subjected to a time-dependent forcing pushing it towards $\mathcal{B}$ (as is the case with $S_{S,f}(x,t)$), the TAMS ensemble exhibits a mean drift towards $\mathcal{B}$ over time.
This results in branching trajectories consecutively closer to $T_a$, and can lead to \emph{late extinction}.
In this case, the ensemble maximum gets closer to $T_a$ with each TAMS iteration and it becomes impossible for the system to reach new maxima.
Although the time-dependence introduced in Eq.~(\ref{eq:dynamic}) partially alleviates the issue, this drift must be accounted for when constructing the data-driven score function.

To that end, we perform deterministic simulations of the system (similar to the ones shown in Fig.~\ref{fig:protocol_deterministic}, with $\alpha_0 < 0.612$), and record the system states. Then, we use these states to define a moving reference state $\mathcal{A}(t)$ in Eqs.~\eqref{eq:sf_naive}, \eqref{eq:sf_baars}, and \eqref{eq:sf_pod}, effectively removing the drift from the computation of the score of the TAMS ensemble.
In the Boussinesq model, $\mathcal{B}$ is only marginally affected by the time-dependent forcing in the considered range of $\alpha_0$ (in Fig.~\ref{fig:protocol_deterministic}, all transitioning trajectories have a static score function $\xi^0_1(\mathbf{X}_t)$ value close to one in the AMOC-off state), such that $\mathcal{B}$ is kept fixed, but the same process could be used to define a moving target state $\mathcal{B}(t)$.

\subsection{Software implementation}
Simulations are performed using the pyTAMS package \citep{Esclapez2025}, a Python implementation of the TAMS algorithm amendable to high-dimensional dynamical systems.
The 2D Boussinesq model is readily available in pyTAMS.
A $41 \times 81$ grid resolution is employed, similar to~\citep{Soons2025}, with fixed step size of $\Delta_t = 0.01$, such that a trajectory run until $T_a=40$ comprises 4000 time steps.
The resulting memory requirements (considering a double precision representation of the four state variables) amounts to $\sim 0.4$ GiB per trajectory.
To mitigate the memory bottleneck of appying TAMS to such a model, pyTAMS only tracks a single, constantly updated state per trajectory, and trajectories are subsampled before being written to disk. In practice, we record the noise increments at every step but store the model state only every fifty steps.
This subsampling leads to a small increase in the computational cost, since part of a trajectory may need to be recomputed during the selection/mutation process, but reduces both the memory requirement and the time spent on input-output operations.

\section{Results}
\label{sec:results}

The TAMS results presented in this section were obtained with a relatively small ensemble size $N = 25$, discarding a single level $Q^{(i)}$ at each iteration of the algorithm ($k_j = 1$). Iterations continued until either all trajectories reached the AMOC-off state, the algorithm reached extinction, or $J=5000$ selection/mutation events were completed (the latter limit was never encountered in the present experiments). 
Unless specified otherwise, the algorithm is repeated $K = 100$ times.

\subsection{Transitions under autonomous forcing}
\label{ssec:auto_results}

The system is first studied in an autonomous forcing configuration ($\alpha_0 = 0$).
We consider noise levels $\epsilon$ between $0.05$ and $0.0125$, resulting in estimated probabilities $\overline{P}_K$ of an AMOC collapse onset before $T_a$ that range from infrequent ($\sim 10^{-2}$) to very rare ($\sim 10^{-8}$).

\subsubsection{Comparison of score functions}
\begin{figure*}[]
\centering
\includegraphics[width=\linewidth]{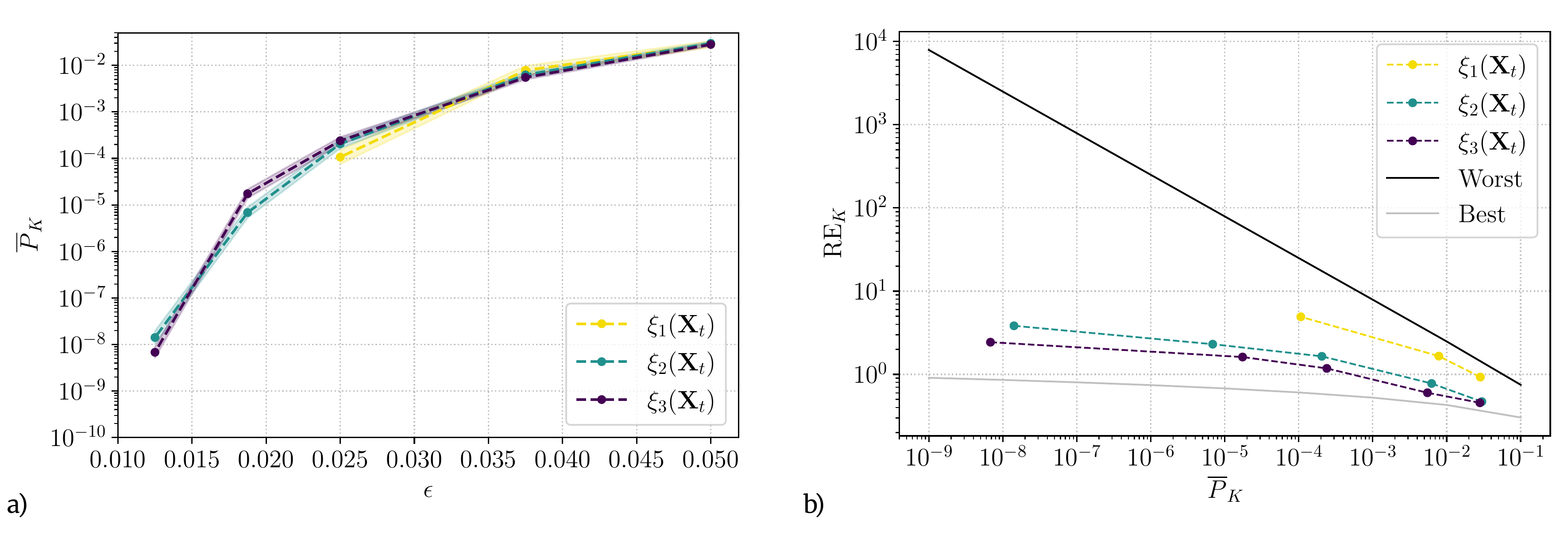}
\caption{(a): Transition probability $\overline{P}_K$ as a function of the noise amplitude $\epsilon$ for the three score functions $\xi_k$. The shaded area corresponds to the 95\% confidence interval $[\overline{P}_K / \Delta_{log} P_K;\overline{P} \times \Delta_{log} P_K]$ with $\Delta_{log} P_K = 1.96 \sigma_{log}/\sqrt{K}$. (b): Relative error $\mathrm{RE}_K$ as a function of the transition probability $\overline{P}_K$, obtained with the three score functions, along with its theoretical lower and upper bounds~\citep{Cerou2019a}.}
\label{fig:autonomous_probability}
\end{figure*}

We estimate $\overline{P}_K$ and its variance using each of the three score functions $\xi_k$ (Fig.~\ref{fig:autonomous_probability}). To construct a meaningful confidence interval (CI) for rare events, we use a 95\% log-normal CI by computing $\Delta_{log} P_K = 1.96 \sigma_{log}/\sqrt{K}$ with $\sigma_{log} = \sqrt{\ln(1+\mathrm{RE}_K^2)}$.
When using the naive score function $\xi_1$, all $K=100$ TAMS runs exhibit extinction when the noise amplitude drops below $0.025$, revealing the poor quality of the score function.
Comparing the relative error $\mathrm{RE}_K$ (see Sect.~\ref{ssec:tams}) of the three score functions with its theoretically worst and best scaling behaviour~\citep{Cerou2019a}, we find that
$\xi_1$ performs significantly worse than the other two score functions, even for high transition probabilities, and scaling closer to the worst case scenario as the transition probability decreases.
The data-driven score function performs best across the entire range of tested noise levels; nonetheless, it still deviates significantly from the best scaling scenario obtained when using the committor function.

\subsubsection{Sensitivity to the number of TAMS runs}
\begin{figure*}[]
\centering
\includegraphics[width=\linewidth]{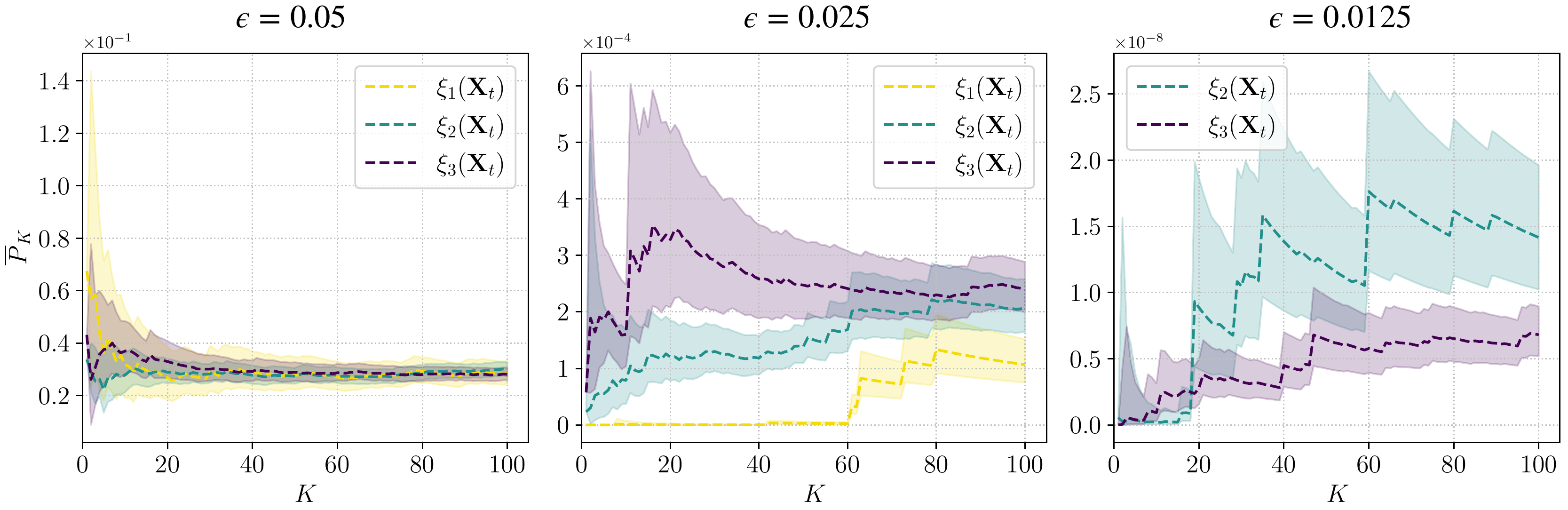}
\caption{Transition probability $\overline{P}_K$ as a function of the number of TAMS realization $K$, at three noise levels $\epsilon\in(0.05, 0.025, 0.0125)$, from top to bottom.}
\label{fig:autonomous_proba_hist}
\end{figure*}

Figure~\ref{fig:autonomous_proba_hist} visualises the convergence behaviour of $\overline{P}_K$ and its CI as $K$ increases, for three different noise amplitudes and all score functions.
At high noise level, all three CIs overlap and decrease smoothly as $K$ increases, suggesting that all three estimators are performant~\citep{Brehier2016}.
As the transition probability decreases with the noise amplitude, the CI overlap is no longer guaranteed and the CIs exhibit large jumps, even after dozens of TAMS realisations, especially for the lowest noise of $\epsilon = 0.0125$.
This behaviour is due to extinctions, which leads to an underestimation of the transition probability~\citep{Brehier2016}.
As the number of TAMS runs increases, this bias is offset by a few runs that largely overestimate the transition probability, such that $\overline P_K$ converges to the true, unbiased value for large $K$.
For example, for $\xi_1$ at intermediate noise ($\epsilon=0.025$), the value of $\overline{P}_K$ exhibits a small initial increase followed by a constant near-zero value up to $K=60$. At this point, a couple of TAMS runs eventually succeed with a high probability, leading to jumps in $\overline{P}_K$ and widening of the CI.
Overall, the behaviour of the data-driven score function $\xi_3$ appears smoother, but increasing $K$ might be necessary to better estimate the transition probability when the transition becomes extremely rare.

\subsubsection{Evolution of the maximum score with TAMS iterations}
\begin{figure*}[]
\centering
\includegraphics[width=\linewidth]{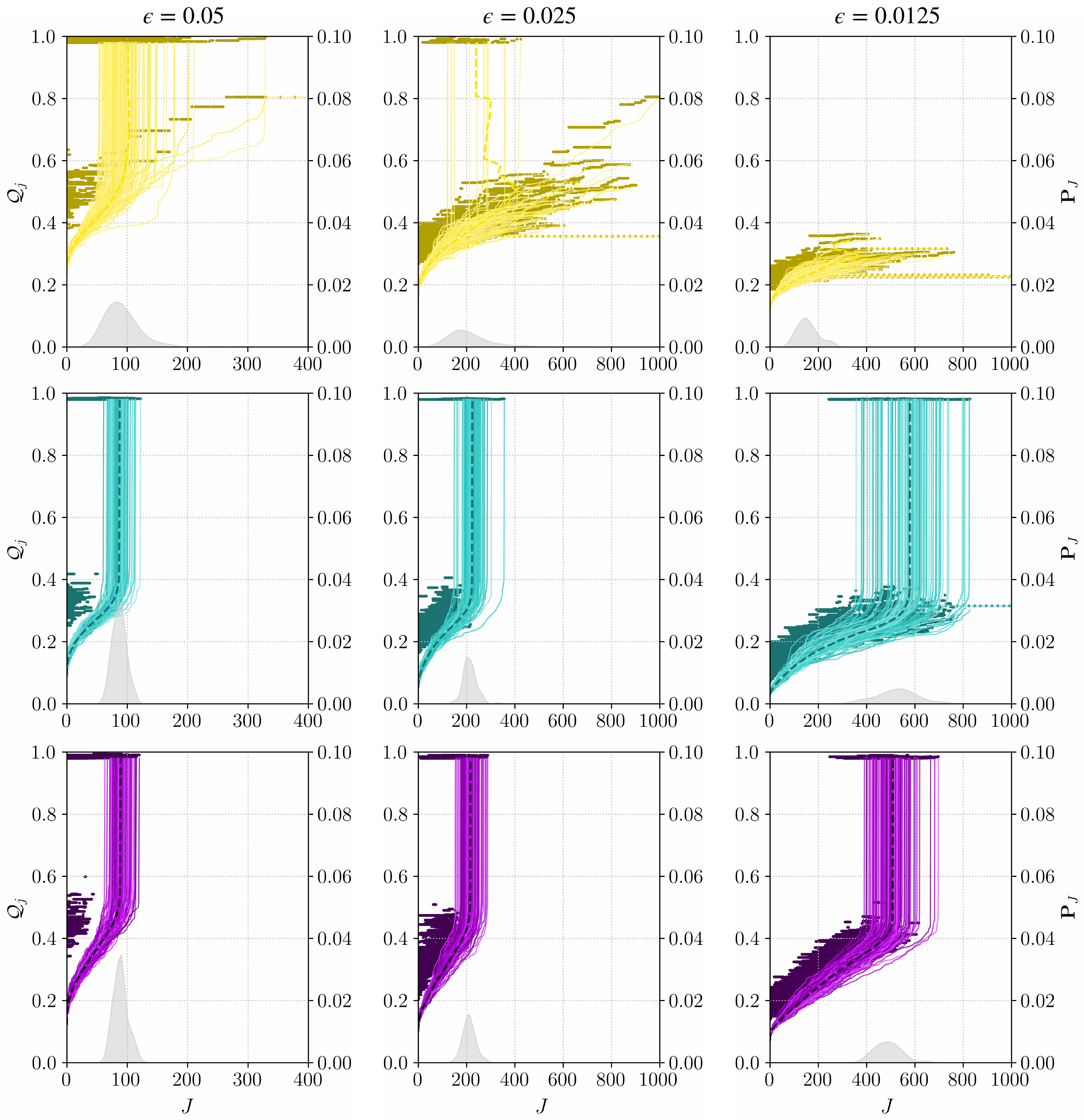}
\caption{Collection of $\mathrm{min}\{Q^{(i)}\}^j_{i\in[1,N]}$ (lines) and $\mathrm{max}\{Q^{(i)}\}^j_{i\in[1,N]}$ (dark dots) across all TAMS runs as a function of the number of selection/mutation events $J$, for each score function $\xi_k(\mathbf{X}_t,t)$ with $k=1,2,3$ (rows/colours, from top top bottom) and noise level $\epsilon = 0.05, 0.025, 0.0125$ (columns). Each coloured line corresponds to a TAMS realization; dashed lines indicate the average number of selection/mutation events $J$ (over the 100 TAMS runs) at which a given level $\mathcal{Q}^j$ is reached. The distribution of $J$-values at which the algorithm stopped is shown (grey shading) with respect to the right y-axes.}
\label{fig:autonomous_minmax}
\end{figure*}

To further analyze the behaviour of the ensemble simulation and the impact of the choice of score function on the ensemble bias, we investigate how the maximum score evolves with TAMS number of selection/mutation events $J$ (Fig.~\ref{fig:autonomous_minmax}).
Results are aggregated across all TAMS realisations obtained with each score function for the three noise levels considered above. For each realisation, we track both the lowest ($\mathcal{Q}_j = \mathrm{min}\{Q^{(i)}\}^j_{i\in[1,N]}$) and highest ($\mathrm{max}\{Q^{(i)}\}^j_{i\in[1,N]}$) maximum score in the ensemble at each algorithm iteration $j$.
Extinctions can be seen via $\mathcal{Q}^j$-lines plateauing below 1 and eventually merging with the $\mathrm{max}\{Q^{(i)}\}^j_{i\in[1,N]}$ value, as observed when using the naive score function $\xi_1$.

In contrast with $\xi_1$, the spread of the distribution of $\mathbf{P}_{J}$ is significantly smaller when using $\xi_2$ or $\xi_3$, consistent with the narrower CI (see Fig.~\ref{fig:autonomous_proba_hist}).
At the lowest noise level, we observe that using $\xi_2$ leads to more outliers compared to $\xi_3$, especially in early iterations, and occasional extinctions are also observed.
Additionally, the lines of $\mathcal{Q}^j$ become less flat using $\xi_3$, indicating that the algorithm is able to consistently bias the ensemble without encountering strong barriers.

All runs exhibit sharp jumps in $\mathcal{Q}^j$ at the last iteration before the algorithm stops and few intermediate values of $\mathrm{max}\{Q^{(i)}\}^j_{i\in[1,N]}$ (between $0.5$ and $1$).
This indicates a labelling mismatch between the score function and committor isolevels.
A threshold $z_{th}$ appears to exist in score function values (closely related to the score level at the edge state) beyond which a given trajectory is virtually guaranteed to transition.
In other words, after passing the edge state, trajectories are bound to transition, meaning that the committor (for our specific problem) at this point is effectively 1, whereas the score function value there is much lower.
For $\xi_2$ and $\xi_3$, $z_{th}$ is close to 0.4 and 0.5, respectively.
Such a labelling mismatch is not necessarily an issue for TAMS, as the algorithm is only sensitive to the ordering and shape of the score function isolevels, not their absolute value.
When using $\xi_1$ a threshold is not as clear and intermediate values of $\mathrm{max}\{Q^{(i)}\}^j_{i\in[1,N]}$ are observed, without necessarily leading to a transition.
This misalignment between the score function and committor isolevels is a common sign of a poor score function.

To quantify the performance of the three score functions with increasing TAMS iterations, we can estimate how well their behaviour reproduces that of the committor.
TAMS effectively splits the rare event probability $\hat{p}$ into a sequence of intermediate, larger, conditional probabilities $p_j$.
Equation~\eqref{eq:ams_estimator} can be rewritten as a function of the intermediate conditional probabilities:
\begin{equation}
    \mathbb{P}(\tau_{\mathcal{B}} < T_a) = \prod_{j=1}^{\tilde{J}} p_j,
\end{equation}
with
\begin{equation}
    p_j = \mathbb{P}\left( \tau_{z_j} < T_a  \mid \tau_{z_{j-1}} < T_a \right),
\end{equation}
for a sequence of score function levels $z_j > z_{j-1}$ adaptively determined by the algorithm, such that $z_{\tilde{J}} = z_{max}$.
While running TAMS, the conditional probabilities are estimated by the survival ratio ($1-l_j/N$), which is the number of conserved trajectories at each iteration.
Let us now derive what it would mean for $p_j$ to use the committor function $q$ as score function.
By definition, $q(z_j)$ gives the probability of reaching $\mathcal{B}$ before $T_a$, starting from any point having a score $z_j$.
But if the committor is used as score, the levels $z_j$ correspond to isolevels of the committor, so the level $z_j$ is the probability $q(z_j)$ itself.
This can be written as:
\begin{equation}
    q(z_j) = z_j = \mathbb{P}\left(\tau_\mathcal{B}<T_a\ |\ \tau_{z_j}<T_a\right).
\end{equation}
Note that the score function levels $z_j$ are strictly increasing as $j$ is increasing.
Therefore, if $z_j$ has been reached before $T_a$, $z_{j-1}$ has also necessarily been reached before $T_a$.
Using Bayes' formula, we can now write:
\begin{equation}
    p_j = \frac{\mathbb{P}\left( \tau_{z_j} < T_a \right)\mathbb{P}\left( \tau_{z_{j-1}} < T_a | \tau_{z_j} < T_a \right)}{\mathbb{P}\left(\tau_{z_{j-1}} < T_a \right)} = \frac{\mathbb{P}\left( \tau_{z_j} < T_a \right)}{\mathbb{P}\left(\tau_{z_{j-1}} < T_a \right)} = \frac{\mathbb{P}\left( \tau_{\mathcal{B}} < T_a \right) / q(z_j) }{\mathbb{P}\left( \tau_{\mathcal{B}} < T_a \right) / q(z_{j-1}) } = \frac{q(z_{j-1})}{q(z_j)} = \frac{z_{j-1}}{z_j}.
\end{equation}

This relation relates the local geometry of the score function ($z_{j-1}/z_j$) to the sequence of conditional probabilities ($1-l_j/N$).
When using the committor, the two are equal on average.
For each of the score functions $\xi_k$, we can thus compute the mismatch between $\overline{r_j} = \langle z_{j-1}/z_j \rangle$ and $\overline{p_j} = \langle 1-l_j/N \rangle$, where the $\langle\cdot\rangle$ operator is an ensemble average over the independent TAMS runs that are not extinct at iteration $j$.
If the mismatch is constant, the score function behaves like the committor but level labels are off (isolevels of the score function are parallel to the committor but offset).

\begin{figure*}[]
\centering
\includegraphics[width=\linewidth]{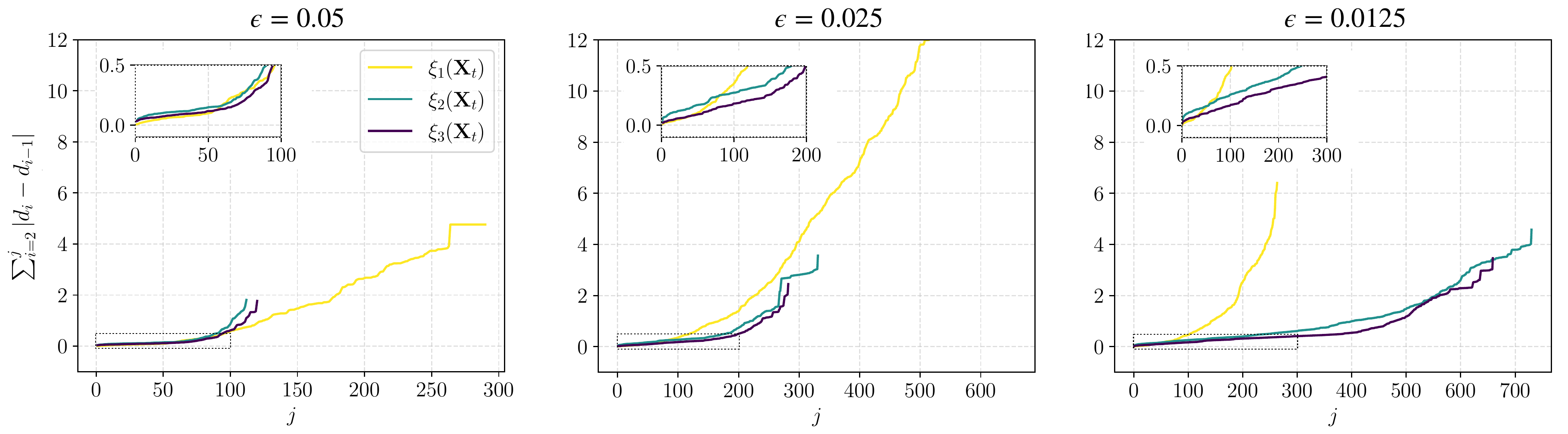}
\caption{Cumulative mismatch $\sum_{i=1}^j |d_{i+1} - d_i|$ as a function of TAMS iteration $j$, averaged over independent TAMS runs for each score function $\xi_k$ (as coloured) and noise level $\epsilon$. Insets magnify the black dotted area in each panel.}
\label{fig:committor_like_behavior}
\end{figure*}

The mismatch between $\overline{r_j}$ and $\overline{p_j}$ is denoted as $d_j = \overline{r_j} - \overline{p_j}$ and accumulates from one iteration to the next (Fig.~\ref{fig:committor_like_behavior}).
Early on, all cumulative mismatch curves slowly increase, corresponding to a regime where the mismatch is dominated by the accumulation of sampling noise.
This is the expected behaviour when using the committor as the score function.
As iterations progress, however, the curves deviate from this behaviour and the mismatch accumulates faster, indicating that the topological mismatch between the committor and the score function now dominates.
This divergence is particularly marked with $\xi_2$ and $\xi_3$, and is related to the jump behaviour of the maximum score (Fig.~\ref{fig:autonomous_minmax}).
However, $\xi_2$ and $\xi_3$ remain in a noise-dominated regime significantly longer than $\xi_1$, especially when the noise decreases.
Additionally, the mismatch for $\xi_3$ is consistently lower than for $\xi_2$, highlighting the smaller volatility.
This measure can give an early-on qualitative indication of the score function quality, by measuring the discrepancy between the score and the committor at every stage of the transition.

Our simulation protocol (cf. Section~\ref{ssec:protocol}) enables us to sample the model state space to determine empirically the position of the basin boundary between the AMOC-on and AMOC-off states while performing TAMS.
We can extract the current model state at time $T_a$ and observe whether a collapse occurs deterministically from that state before $2T_a$.
For this we can use all sampled trajectories, including the ones discarded in TAMS. As shown for the case $\epsilon=0.025$ in different projections of the reduced state space, the separation between the two basins of attraction is clearly visible regardless of the score function, with the two outcomes only overlapping in a narrow region in of the projected phase space.
The estimated basin boundary is close to the edge state (an imperfect intersection is possible due to finite noise) and closely follows the isocontour $\xi_3 = 0.38$.
The $(\mathbf{v}_1,\mathbf{v}_3)$ projection shows the sharpest demarcation between both basins, also captured in the nearly discontinuous score function along the boundary.

\begin{figure*}[]
\centering
\includegraphics[width=\linewidth]{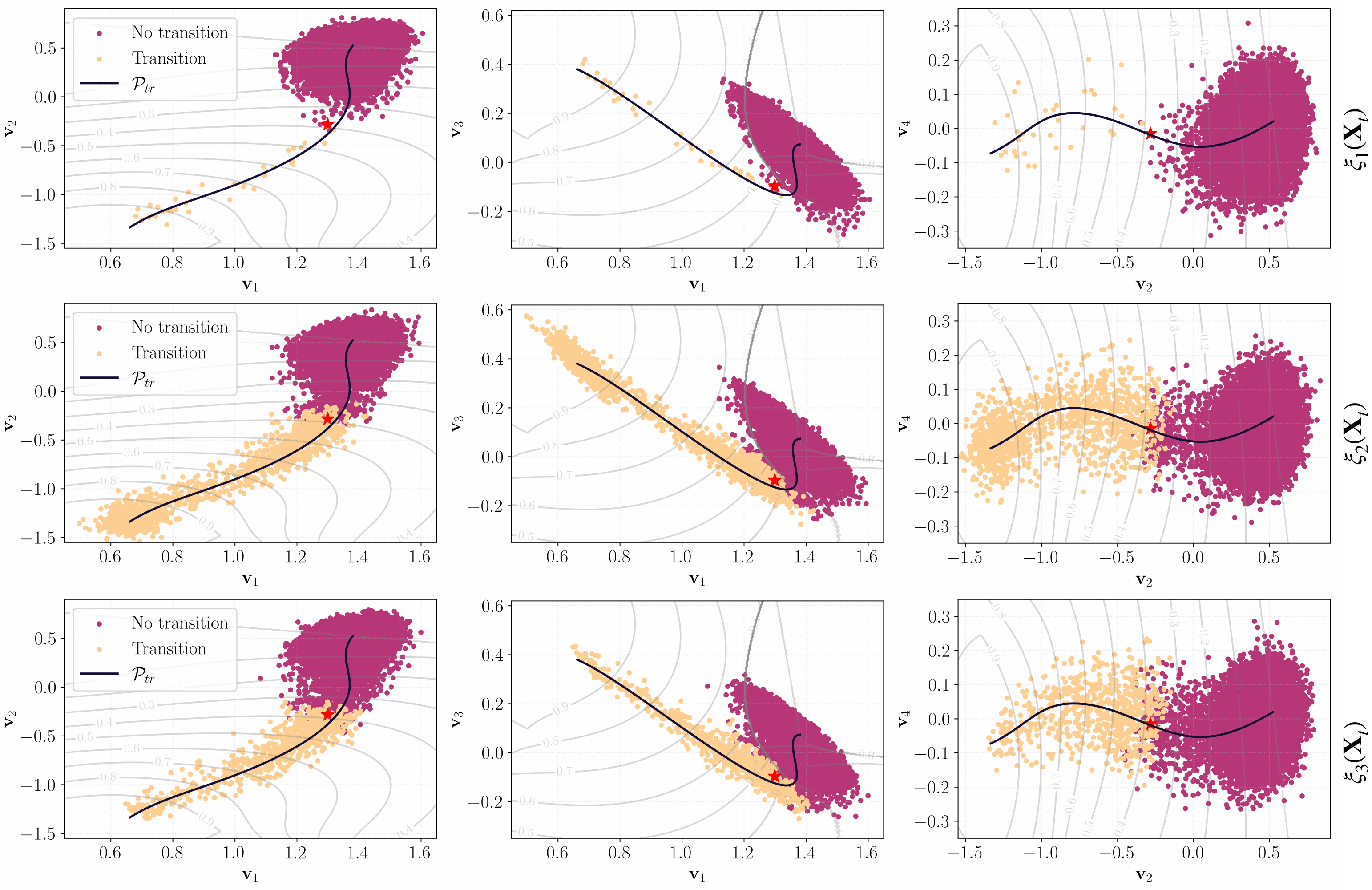}
\caption{Scatter plots of model states at $t=T_a=20$, projected onto the POD space and coloured by the trajectory outcome (reaching AMOC-off by $2T_a$ or not). Data include all trajectories obtained with TAMS runs performed at $\epsilon=0.025$ with score function $\xi_1$ (top), $\xi_2$ (centre) and $\xi_3$ (bottom). The ensemble MTP $\mathcal{P}_{tr}$ (black line), the edge state (red star) and isocontours of $\xi_3$ (grey) are included for reference. Rows correspond to the three score functions, while each column shows a different pairs of the POD space dimensions.}
\label{fig:autonomous_BasinBndy}
\end{figure*}

\subsection{Transitions under time-dependent forcing}

Using the protocol described in Sect.~\ref{ssec:protocol} with time-dependent forcing, we now quantify the AMOC transition probability varying two parameters: the hosing amplitude $\alpha_0$ of the time-dependent forcing at $T_a$ and the noise amplitude $\epsilon$. Due to the destabilising effect of the time-dependent forcing pushing the system closer to the basin boundary, we now use a range of noise amplitudes that is one order of magnitude smaller than the one used under fixed forcing conditions (cf. Sect. \ref{ssec:auto_results}).

\begin{figure*}[]
\centering
\includegraphics[width=\linewidth]{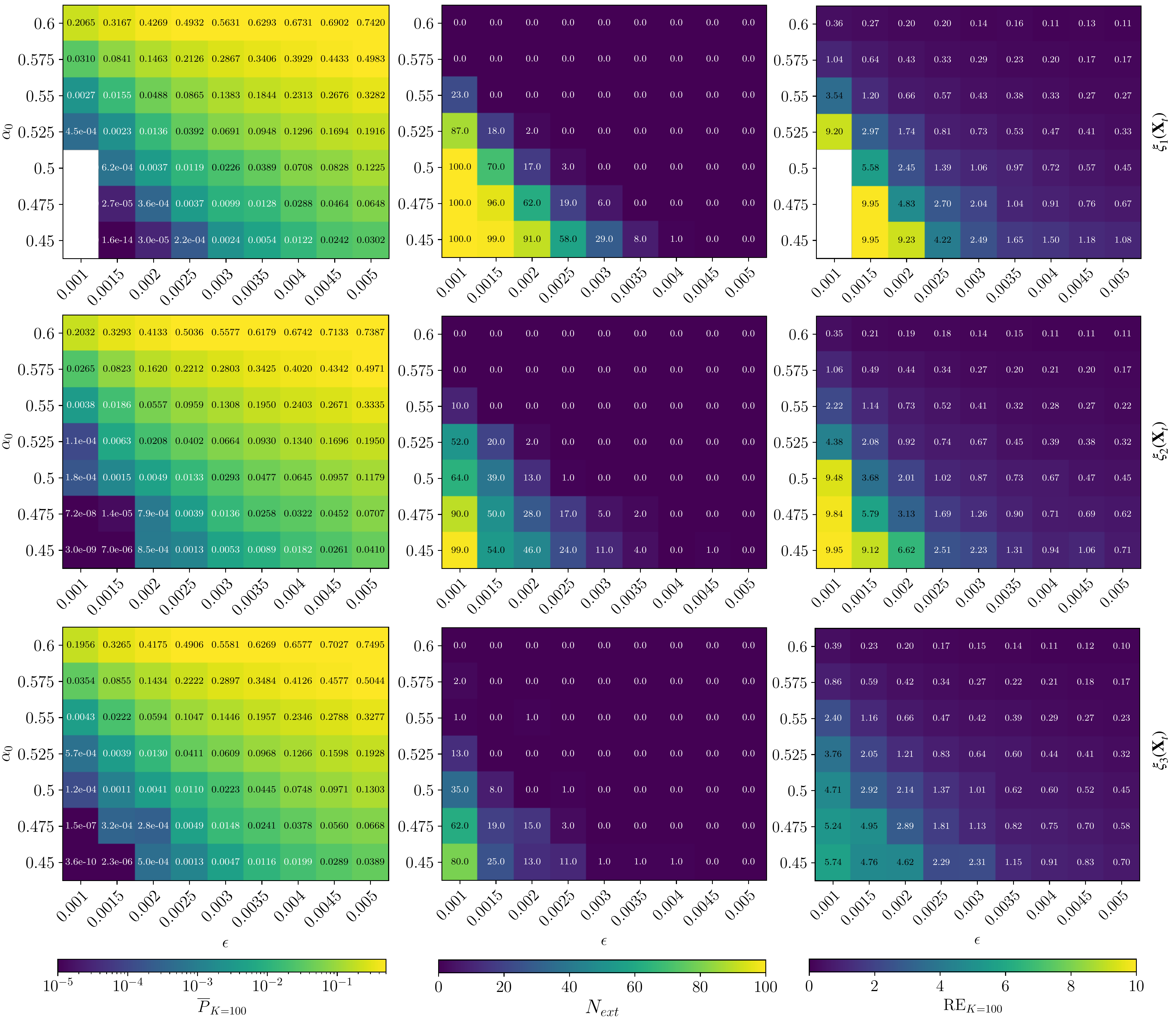}
\caption{Parameter maps (final hosing strength $\alpha_0$ vs. noise strength $\epsilon$) of transition probabilities $\overline{P}_K$ (left), number of extinctions (centre) and transition probability relative error $\mathrm{RE}_K$ (right) using the three score functions $\xi_k(\mathbf{X}_t, t)$ for $k=1,2,3$ (from top to bottom).}
\label{fig:nonauto_allmaps}
\end{figure*}

For each dynamic score function $\xi_k(\mathbf{X}_t,t)$ with $k \in \{1,2,3\}$, we construct 2D maps $(\alpha_0$ vs. $\epsilon$) of the transition probability $\overline P_K$, number of extinctions and the relative error $\mathrm{RE}_K$ estimated from a total of 6300 independent TAMS runs (i.e., $100$ runs of TAMS are performed for each of the nine values of $\epsilon$ and seven values of $\alpha_0$) (Fig.~\ref{fig:nonauto_allmaps}).

Under strong hosing (large $\alpha_0$), the system easily transitions to an AMOC-off state, even with a very low noise amplitude.
The transition probability is affected by the noise level, but the transition is only about four times less likely
at low noise levels than at high noise levels.
As the deterministic forcing $\alpha_0$ decreases, the system transitions become increasingly more difficult to trigger and low noise is now order of magnitudes more unlikely to lead to an AMOC collapse than high noise.

In agreement with the results obtained in the autonomous forcing cases, all three score function coincide very well in the region where $\overline{P}_K > 10^{-2}$, but the results obtained with $\xi_1(\mathbf{X}_t,t)$ degrade when the transition probability drops, with extinction becoming more frequent; eventually, a 100\% extinction is observed at the lowest noise level and low hosing forcing.
The relative error $\mathrm{RE}_{K}$ is found to increase significantly as the transition probability decreases, with $\mathrm{RE}_{K}$ reaching close to 10 for low values of $\alpha_0$ and $\epsilon$.
In this region, extinction is observed with all score functions, albeit significantly less often when using $\xi_3(\mathbf{X}_t,t)$ compared to the other score functions.
The relative error obtained with $\xi_3(\mathbf{X}_t,t)$ is at best two times lower than that obtained with $\xi_2(\mathbf{X}_t,t)$ and both are on par in the high $\overline{P}_K$ region, consistent with the data reported in Fig.~\ref{fig:autonomous_probability}(b).

\begin{figure*}[]
\centering
\includegraphics[width=\linewidth]{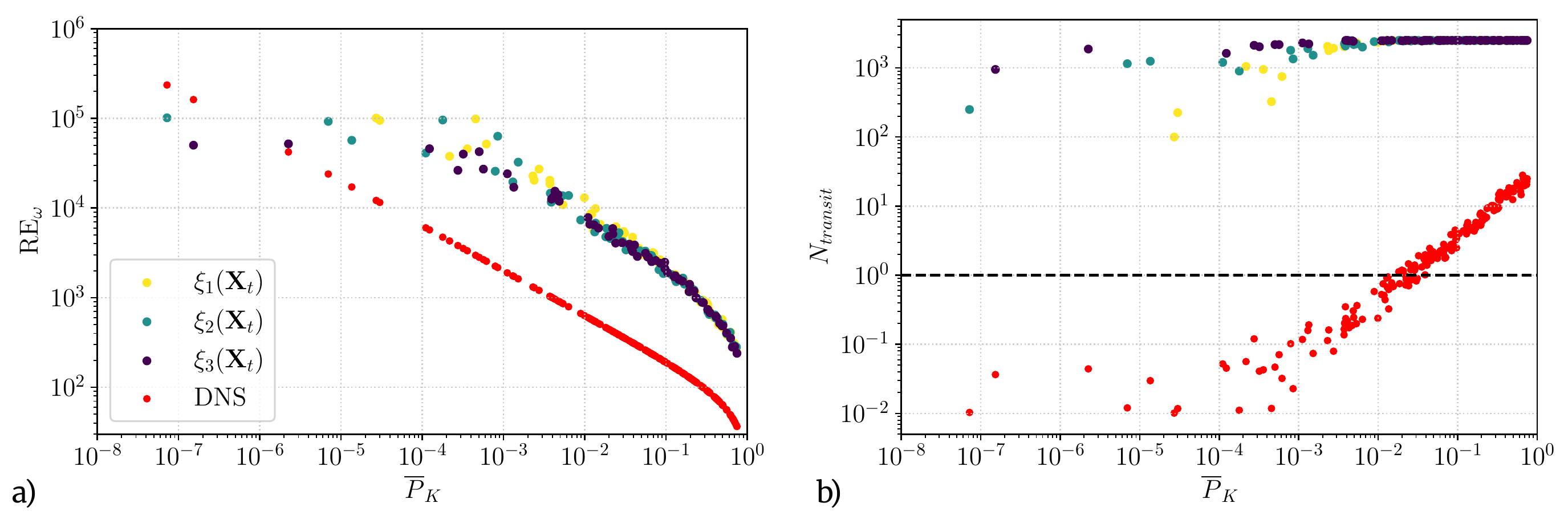}
\caption{(a) Work-normalised relative error $\mathrm{RE}_\omega$ as a function of the transition probability estimate $\overline{P}_K$ for all parameters and dynamic score functions presented in Fig.~\ref{fig:nonauto_allmaps}. Red dots correspond to $\mathrm{RE}_\omega^{DNS}$, the estimated cost of DNS at the same level of relative error as each TAMS data point. (b) Number of transition events $N_{transit}$ obtained running TAMS, or DNS at the same level of relative error. The horizontal dashed line marks the threshold below which DNS might lead to no occurrence of the rare event.}
\label{fig:RE_omega_nonautonomous}
\end{figure*}

The performance of the score functions can also be compared with respect to efficiency by taking into account the computational cost.
We define the work-normalised relative error $\mathrm{RE}_{\omega}$,
\begin{align}
\mathrm{RE}_{\omega} = \frac{\sigma_K}{\overline{P}_K} \times \sqrt{\omega},
\end{align}
where $\omega$ is the computational cost of the TAMS runs.
The total number of model steps is used as a proxy of the computational cost of a TAMS run since the cost of the TAMS algorithm itself (generating noise increments or traversing and sorting trajectories) is negligible compared to advancing the model.
For all three score functions, we find that $\mathrm{RE}_{\omega}$ increases by three orders of magnitude as the transition probability drops from $\sim$ 0.1 to $10^{-7}$ (Fig.~\ref{fig:RE_omega_nonautonomous}a). At the same time, $\mathrm{RE}_K$ only increases by two orders of magnitude, highlighting the increasing cost of TAMS (due to a larger number of iterations and thus of simulated time steps needed) as the target probability decreases.
For high transition probabilities ($\overline P_K \gtrsim 10^{-1}$), all three score functions have similar values of $\mathrm{RE}_{\omega}$.
But as the transition probability decreases, $\xi_1$ shows significantly higher work-normalised relative error, while the difference between $\xi_2$ and $\xi_3$ is close to that observed on $\mathrm{RE}_K$.
Indeed, $\omega$ is found to be fairly similar for $\xi_2$ and $\xi_3$ (not shown here).
An estimate of the relative cost of TAMS against Monte-Carlo can be constructed by computing the number of samples required to reach a relative error equivalent to that of TAMS: $N_{MC} = \frac{1 - \overline{P}_K}{\overline{P}_K \cdot \mathrm{RE}_K^2}$, and assuming that the cost of each sample is $\omega_{0} = 4.0\times10^3$ (i.e the full trajectory length, worst case scenario). Then we can compute $\mathrm{RE}_\omega^{DNS} = \sqrt{N_{MC} \times \omega_{0}} \times \mathrm{RE}_K$. Figure~\ref{fig:RE_omega_nonautonomous}a indicates that TAMS becomes more efficient than DNS only in the low $\overline{P}_K$ regime \citep{Lestang2018}, were improvements brought by the proposed score function are concentrated.
Thus, with regard to maximising the precision of the probability estimate at minimal computational cost, DNS outperforms TAMS for probabilities larger than around $10^{-5}$. However, this neglects the fact that the number of sampled transitions with DNS can be very low, or even zero. When considering the efficiency in terms of the number of sampled transitions at a given relative error and computational cost, TAMS becomes the preferable choice
We can compute the number of transition events obtained with TAMS: $(100-N_{ext})\times25$ and DNS: $N_{MC} \times \overline{P}_K$. Fig.~\ref{fig:RE_omega_nonautonomous}b shows that TAMS enables to capture more than two orders of magnitude more rare events, providing a wealth of data for analysis on top of the probability estimate. Additionally, we see that the average number of rare-events for DNS drops below 1 as the transition probability is lower than $\sim 2\times10^{-2}$, which suggests that more DNS samples would effectively be required to capture a single transition (while running the risk of not getting any occurrence of the rare event at all).

\section{Summary and Discussion}
\label{sec:disc}

A timely issue in predicting future Atlantic Meridional Overturning Circulation (AMOC) behaviour is the estimation of the probability that the onset of a collapse will occur before the year 2100.
Extrapolated statistical early-warning signals, when applied to time series of sea surface temperature-based reconstructions of the AMOC strength, have suggested that the AMOC could start to collapse around mid-century~\citep{Ditlevsen2023}.
Climate model simulations under different greenhouse gas forcing scenarios also show physically derived indications of a collapse onset around that time~\citep{vanWesten2025_JGR, Romanou2023, borner_boundary_2025, Drijfhout2025}.
However, these results do not provide an estimate of the associated transition probability because they are based on single realisations or small initial condition ensembles.

Even for low-dimensional models, the computational cost of Monte Carlo sampling increases drastically with decreasing transition probabilities (e.g., under low noise).
Rare event algorithms, such as Trajectory-Adaptive Multilevel Splitting (TAMS), are crucial for determining probability estimates of climate tipping events with acceptable variance on the estimator. In the context of AMOC transitions,
TAMS has been successfully applied in a low-dimensional ocean model~\citep{Castellana2019} to estimate probabilities down to $10^{-9}$, and in a spatially two-dimensional ocean circulaiton model similar to the version studied here~\citep{Baars2021}. However, owing mainly to an inefficient score function, it turned out challenging to constrain the variance of the estimated transition probability.

Especially in high-dimensional models, the choice of the score function is crucial to limit the computational cost of TAMS and ensure the reliability of the probability estimator.
Here, we addressed the problem of score function design in a latitude-depth Boussinesq model of the thermohaline circulation~\citep{Dijkstra1997,Soons2025} with $\sim 10^4$ degrees of freedom. We considered spatially correlated but temporally white surface freshwater noise, both in stationary forcing conditions and combined with time-dependent deterministic freshwater forcing.
TAMS is well suited for both settings, as it is agnostic to the nature of the imposed forcing.
We presented a novel data-driven score function and tested its performance against two previously proposed candidates.

The naive choice of a score function based on the AMOC strength performed poorly, leading to frequent extinctions (failure to estimate the onset probability) and a large variance of the probability estimator.
By incorporating knowledge of the AMOC equilibrium states, the geometric score function proposed by \citep{Baars2021} significantly reduced the risk of extinction and yielded a considerably smaller variance, especially for low transition probabilities.
The data-driven score function introduced in this paper performed best overall, tackling the high-dimensionality challenge by combining a linear dimension reduction with a non-linear path reconstruction in the latent space. We demonstrated that, compared to the other two score functions, its behaviour is closest to that of the committor, even as the rare event probability decreases.

It appears that the most effective way to improve the efficiency and accuracy of TAMS is to improve its score function, with the optimal score function being the committor function itself.
This calls for the development of techniques for the estimation of the committor, and their coupling with TAMS.
The estimation of the committor, in particular using machine learning, is an active field of research in computational chemistry~\citep{Trizio2025}.
However, the coupling of such data-driven techniques with rare-event algorithms~\citep{Lucente2022,Jacques-Dumas2024} largely remains an open problem.
Our score improvement loop could thus benefit from a combination with data-driven methods more specifically tailored to estimating the committor.

Also, other extreme-event sampling methods; particularly Giardina-Kuchan-Tailleur-Lecomte (GKTL)~\citep{Lestang2018} and Quantile Diffusion Monte-Carlo~\citep{Webber2019} have been used in climate applications.
By design, both are suited for the sampling of events happening before a time horizon, and GKTL has recently been applied to the problem of the AMOC collapse~\citep{Cini2024}.
Despite their lower computational cost, these algorithms are not as well adapted as TAMS for the estimation of transition probabilities.
On the methodological level, it may be possible to design hybrid methods that can leverage the strengths and weaknesses of these different algorithms.
Along these lines, \citep{Chraibi2021} hinted at a possible improvement of the estimator variance via a probabilistic discarding step in TAMS.
However, this modification of the algorithm also relies on knowledge of the committor, which appears once again to be the main bottleneck for improvement.

In this paper, we presented a test case of applying TAMS to high-dimensional ocean and climate models, with the goal of estimating the probability of an AMOC collapse onset given a time horizon and forcing protocol. While we used a simplified Boussinesq model to be able to compare different score functions and parameter settings, its spatiotemporal dynamics offer methodological lessons transferable to three-dimensional global ocean models. As a next step, we aim to leverage these insights to explore rare events in the Parallel Ocean Program (POP) ocean model, driven by anthropogenic climate change forcing and stochastic surface flux variability~\citep{Boot2025_Noise}. Although this involves a system size increase by two orders of magnitude (compared to the Boussinesq model), the relatively small ensemble size selected here is also in the feasible realm for POP. An additional challenge will be that the stability properties (e.g. location of the edge state and AMOC-off state) and candidate transition paths (as given here by the instanton) are less known in these large models. Nonetheless, we argue that the use of TAMS, if appropriately configured, is technically possible in Earth System Models -- bringing reliable estimates of AMOC transition probabilities within reach.

\section*{Acknowledgements}
The work of LE, VJD and HD was funded by the European Research Council through the ERC-AdG project TAOC (project 101055096, PI:Dijkstra).
The work of RB was supported by the ClimTip project, which has received funding from the European Union's Horizon Europe research and innovation programme under grant agreement No. 101137601.
LE and LS were also funded by the Netherlands eScience Center, through the eTAOC project.
The model simulations were conducted on the Dutch National Supercomputer Snellius within NWO-SURF project 2024.13.
We thank Jelle Soons (IMAU) for important discussions on the instanton of the Boussinesq model.

\section*{Conflict of Interest}
The authors have no conflicts of interest to disclose.

\section*{Author Contributions}
{\bf Lucas Esclapez}: Investigation, Visualization, Methodology, Conceptualization, Software, Formal analysis, Writing - original draft, Writing - review and editing.
{\bf Val\'erian Jacques-Dumas}: Methodology, Conceptualization, Software, Formal analysis, Writing - original draft, Writing - review and editing.
{\bf Reyk B\"orner}: Conceptualization, Software, Formal analysis, Writing - review and editing.
{\bf Laurent Soucasse}: Methodology, Software, Writing - review and editing.
{\bf Henk A. Dijkstra}: Conceptualization, Resources, Funding acquisition, Writing - review and editing.

\section*{Data Availability Statement}
The Python implementation of the Boussinesq model is available online \citep{Esclapez2025}.
The model outputs and the scripts used to produce the figures presented in the manuscript are openly available at \url{https://doi.org/10.5281/zenodo.18923963} \citep{Esclapez:2026}.

\appendix
\numberwithin{equation}{section}

\section{Variations of the data-driven score functions}
\label{sec:data_driven_tests}

\begin{figure*}[]
\centering
\includegraphics[width=\linewidth]{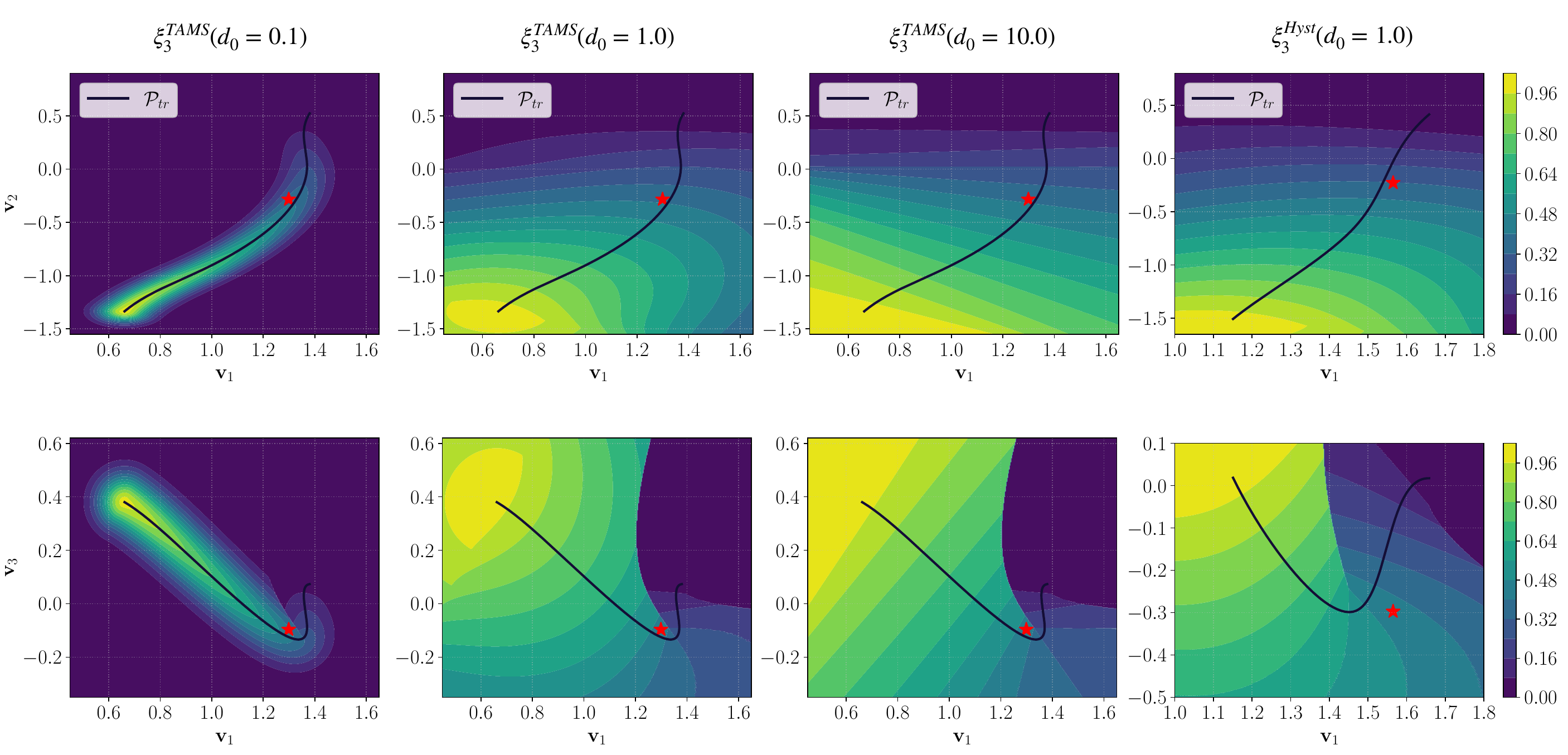}
\caption{From left to right: data-driven scores $\xi_3^{TAMS}$ with $d_0=0.1$, $d_0=1.0$ and $d_0=10$, as well as $\xi_3^{Hyst}$. Each score is plotted in the projection $\mathbf{v}_1-\mathbf{v}_2$ (top row) and $\mathbf{v}_1-\mathbf{v}_3$ (bottom row) of the POD latent space. The red star indicates the position of the edge state.}
\label{fig:var_score_map}
\end{figure*}

The results presented with the data-driven score function $\xi_3(\mathbf{X}_t,t)$ in Sect.~\ref{sec:results} were all obtained using a POD latent space and a MTP based on TAMS data, and a constant value of $d_0=1.0$.
Two variations are explored here:
\begin{enumerate}
    \item Using the data from the upper branch of a quasi-steady hysteresis experiment to construct the POD latent space and MTP (effectively the bifurcation path). The resulting score is called $\xi_3^{Hyst}$.
    \item Increasing or decreasing $d_0$ by one order of magnitude ($d_0 = 0.1$ and $d_0 = 10.0$), while using the same POD space and MTP as in Sect.~\ref{sec:results}. The resulting score is called $\xi_3^{TAMS}$.
\end{enumerate}
Figure~\ref{fig:var_score_map} shows both scores plotted against the main POD modes latent space.
Its first three columns correspond $\xi_3^{TAMS}$ computed with the three values of $d_0$, while its fourth column presents $\xi_3^{Hyst}$.
Using a value of $d_0$ much smaller than the range of the MTP in the latent space leads to a fast decay of the score function, effectively constraining transitions through a narrow tube around the MTP.
As $d_0$ increases, the isolevels of the score function extend further away in the directions transverse to the MTP.
Even though the score function produced when using the hysteresis data appears qualitatively close to that obtained using TAMS data and $d_0=1.0$ (second column of Fig.~\ref{fig:var_score_map}), the MTP differs enough to pass further away from the edge state (especially on higher POD modes, see the bottom right panel of Fig.~\ref{fig:var_score_map}).

To provide a quantitative comparison, tests are performed in the autonomous forcing case with $\epsilon = 0.0125$ (see Sect.~\ref{ssec:auto_results}), and compared to the reference results obtained using $\xi_3^{TAMS}$ with $d_0=1.0$.
Table~\ref{tab:xi3_var_results} compares the four variations of the data-driven score function in terms of mean transition probability $P_K$, relative error $\mathrm{RE}_K$, number of observed extinctions $N_{ext}$ and average computational cost $\omega$.
Using $\xi_3^{TAMS}(d_0=0.1)$ leads to a significantly lower transition probability and large relative error, due to the large number of extinctions observed.
By constraining the meaningful score information close to the MTP, a small $d_0$ does not allow the system to explore the phase space under the effect of the noise in order to find other transition paths.
The three other variations are close together in terms of transition probability and relative error, even though the hysteresis-based score shows a higher number of extinctions and an associated increase in relative error.
A closer look at the data sampled from the TAMS runs using the hysteresis-based score shows that extinctions occur close to the AMOC-off state.
This is due to the difference between the AMOC-off state obtained without deterministic hosing ($\alpha(t) = 0.0$) and the one reached through the hysteresis experiment ($\alpha(t) \neq 0.0$), corresponding to a bifurcation-induced transition.
This is indeed a concern since there are no guarantees that a noise-induced transition would proceed through the same path as a bifurcation-induced transition.

\begin{table}[h]
\centering
\begin{tabular}{| l | c | c | c | c |}
\hline
\textbf{Score function} & $\overline{P}_K$ & $\mathrm{RE}_K$ & $N_{ext}$ & $\omega$ \\ [1pt] \hline
$\xi_3^{TAMS}(d_0=0.1)$ & $6.55\times10^{-10}$ & 7.71 & 43 & $1.71\times10^{6}$ \\ [1pt] \hline
$\xi_3^{TAMS}(d_0=1.0)$&  $6.81\times10^{-9}$ & 2.43 & 0 & $1.59\times10^{6}$ \\ [1pt] \hline
$\xi_3^{TAMS}(d_0=10.0)$& $9.98\times10^{-9}$ & 2.66 & 1 & $1.59\times10^{6}$ \\ [1pt]\hline
$\xi_3^{Hyst}(d_0=1.0)$& $7.43\times10^{-9}$ & 3.37 & 19 & $2.04\times10^{6}$ \\ [1pt] \hline
\end{tabular}
\caption{Compared performances of the four data-driven score functions tested here under autonomous forcing at $\epsilon = 0.0125$.}
\label{tab:xi3_var_results}
\end{table}

\begin{figure*}[]
\centering
\includegraphics[width=0.8\linewidth]{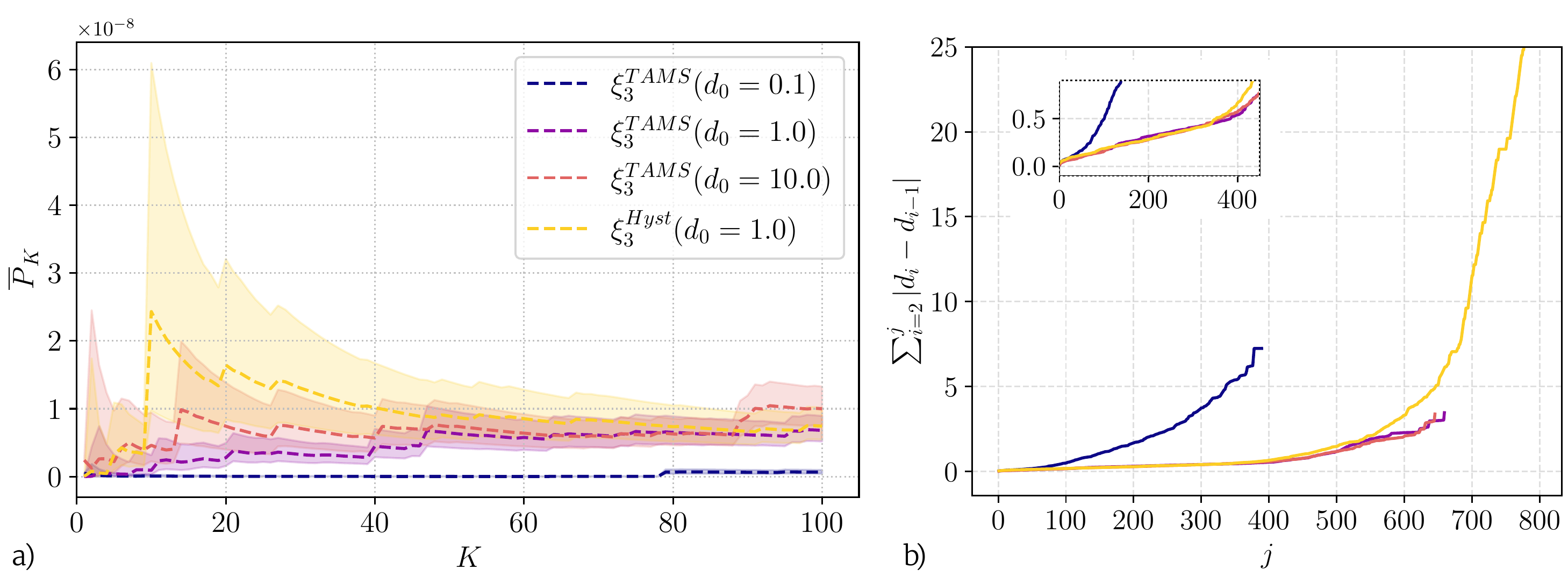}
\caption{(a): Evolution of $\overline{P}_K$ as a function of the number of TAMS realisations $K$. (b): Cumulative absolute variation in the mismatch $d_j = \overline{r_j} - \overline{p_j}$ between $\overline{r_j} = \langle z_{j-1}/z_j \rangle$ and $\overline{p_j} = \langle 1-l_j/N \rangle$ over the course of TAMS iterations.}
\label{fig:var_score_perfs}
\end{figure*}

To further evaluate the behaviour of the TAMS estimator and the TAMS ensemble biasing, we provide in Fig.~\ref{fig:var_score_perfs} the evolution of $\overline{P}_K$ as a function of the number of independent TAMS realizations $K$ and the measure of the discrepancy between each score function tested here and the committor function (see Fig.~\ref{fig:committor_like_behavior} and discussion).
Both graphs confirm that $\xi_3^{TAMS}(d_0=0.1)$ is particularly ill-behaved.
The confidence intervals of the three other score functions are overlapping almost through the entire range of $K$ values considered.
Those three variations also exhibit committor-like behaviour during the first 350 TAMS iterations, after what $\xi_3^{Hyst}$ increases faster than the other two, which remain close to each other through the entire iteration process.

Overall, the choice of $d_0$ does have an impact on the performance of the data-driven score function proposed in this work.
However, our tests indicate that the performance is only marginally affected if $d_0$ is of the same order of magnitude as the largest span of the MTP in the latent space.
Future improvements could focus on using a non-homogeneous value of the $d_0$ parameter (i.e. using a different $d_0$ value in each dimension of the latent space).
Using a quasi-steady hysteresis experiment dataset instead of TAMS run data to build the score function does not lead to a poor score function for the present system (compared to the naive score function tested in the main text), but there is no guarantee that noise-induced and bifurcation-induced paths are as close for other systems.

\section{Test of the data-driven score function on a simple model}
\label{sec:test_datascore}
In order to evaluate the performance and applicability of the data-driven score function formulation to other systems, we study the case of a 2D Langevin dynamics in a triple-well potential \citep{Rolland2015, Brehier2016}.
The system features two symmetric global minima and one local minimum, allowing for two distinct reactive channels, depending on the noise applied to the system.
The system dynamics are given by:
\begin{align}
\frac{\mathrm{d}x}{\mathrm{d}t} &= - \frac{\partial V}{\partial x} + \sqrt{\frac{2}{\beta}} \mathrm{d} W \\
\frac{\mathrm{d}y}{\mathrm{d}t} &= - \frac{\partial V}{\partial y} + \sqrt{\frac{2}{\beta}} \mathrm{d} W,
\end{align}
where $\beta$ is the inverse temperature, effectively controlling the noise in the system and $V$ is the potential given by:
\begin{equation}
V(x,y) = \frac{x^4}{5} +\frac{1}{5}\left( y-\frac{1}{3}\right)^2+3e^{-x^2-(y-1/3)^2}-3e^{-x^2-(y-5/3)^2} - 5e^{-(x-1)^2-y^2} - 5e^{-(x+1)^2-y^2}.
\end{equation}

\begin{figure*}[]
\centering
\includegraphics[width=\linewidth]{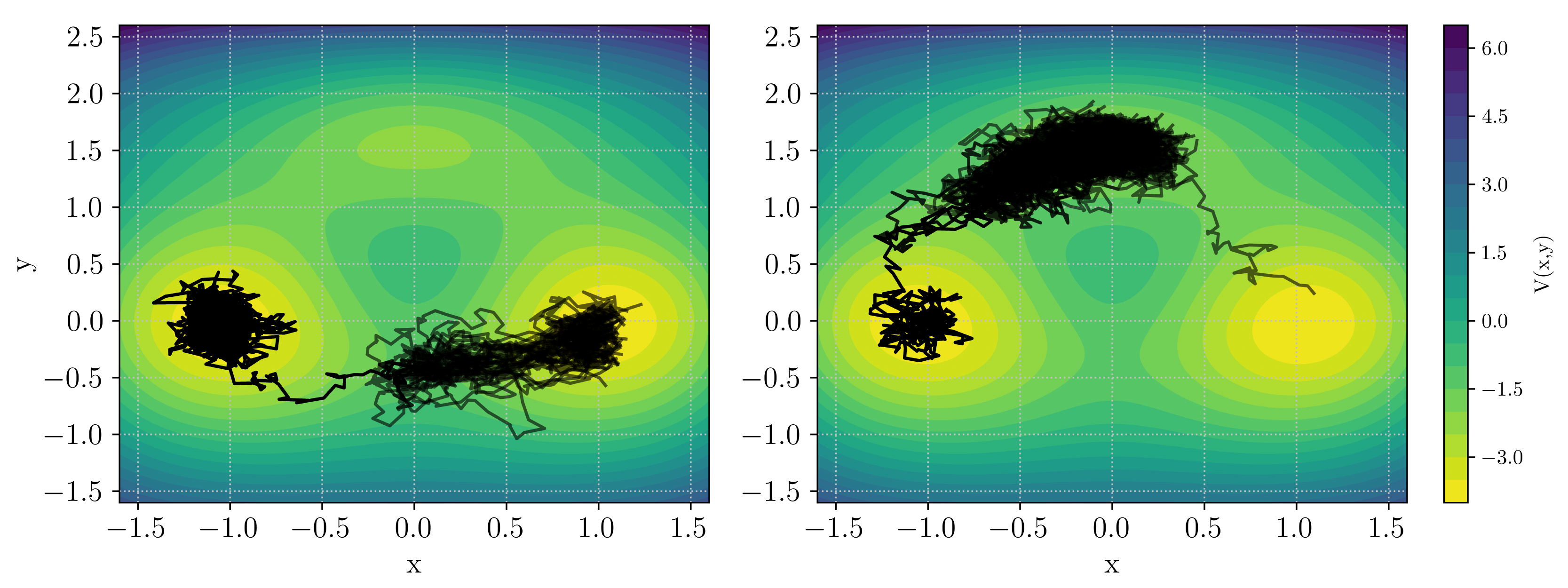}
\caption{2D triple-well potential with reactive trajectory through the lower channel (left) and the upper channel (right).}
\label{fig:potential_triplewell}
\end{figure*}

As shown in Fig.~\ref{fig:potential_triplewell}, the global minima are located at $(\pm1,0)$ and the local minimum is located at $(0,1.5)$, providing an example of transition paths through the upper and lower channels.
We are interested in transitions from the left global minimum to the right one.
At high temperature (small $\beta$), reactive paths go through the lower channel, as the large noise allows overcoming the high energy barrier.
At low temperature (large $\beta$), transition paths will favor the upper channel, going through the local minimum and crossing two lower energy barriers.

The data-driven score function is built using the process described in Sect.~\ref{ssec:data-driven-score}, without the use of POD, as the system only features two dimensions.
We select a fixed value of $\beta$ at $4.67$, near the cross-over between the lower and upper channels phase transition~\citep{Rolland2015}, such that transitions through both channels are expected.
At first, a set of ten TAMS run is performed with $N=32$, $T_a = 20$ and a simple score function solely based on the $x$ coordinate:
\begin{equation}
\xi_1(x,y) = \frac{x+1}{2}.
\end{equation}

\begin{figure*}[]
\centering
\includegraphics[width=\linewidth]{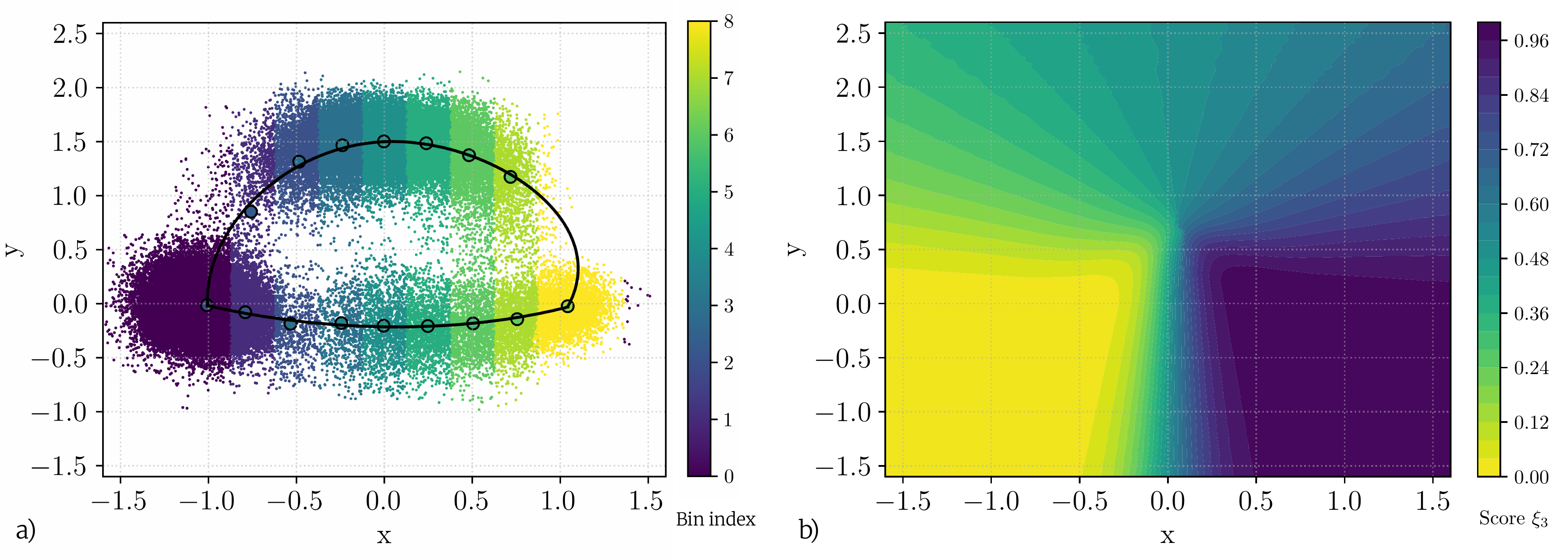}
\caption{a) Model states along transitioning trajectories obtained from ten TAMS runs (coloured scatterplot), along the reconstructed MTPs (lines) obtained from the paths barycentres (large circled dots). b) Score function $\xi_3(x,y)$ obtained from Eq.~(\ref{eq:sf_pod}).}
\label{fig:triplewell_firstpass}
\end{figure*}

The model states along the transition paths obtained with TAMS, as well as the two MTPs constructed by the algorithm are presented in Fig.~\ref{fig:triplewell_firstpass}(a).
Data points are coloured by the index of the bin they fall into (nine bins were used here), highlighting the simplicity of the initial score function.
Fig.~\ref{fig:triplewell_firstpass}(b) shows the data-driven score $\xi_3(x,y)$ based on the distance-weighted average of both paths.

We then iterate the method through ten additional runs of TAMS, driven this time by the score function $\xi_3(x,y)$ depicted in Fig~\ref{fig:triplewell_firstpass}(b).
Only minor changes to the score map are observed with this new data set, indicating that $\xi_3(x,y)$ is fairly robust to the input dataset, as long as transitions along both channels are present in the set.
The score function shown in Fig~\ref{fig:triplewell_firstpass}(b) captures the main features of the committor fonction for this specific problem (see~\citep{Rolland2015} for instance).
The most notable difference is observed in the upper channel, where the committor remains flat across the entire local minimum $(0.0,1.5)$ basin, with clear narrow strong gradients near its left and right boundaries, whereas the present score function exhibits a smooth transition along the upper MTP.
Further improvement could therefore be considered, using for instance the input data local density.
But such approach would not necessary be applicable to our target larger dynamical systems, where not as much data is available.

\section{Mean transition path vs. instanton}
\label{sec:instanton_vs_mtp}

As discussed in Sect.~\ref{ssec:data-driven-score}, the mean transition path (MTP) obtained in the construction of the data-driven score function differs from the Freidlin-Wentzell instanton computed by~\citep{Soons2025} (see Fig.~\ref{fig:ensemblepath_inPOD}).
Specifically, the MTP deviates from the instanton in the segment between the AMOC-on state and the edge state, whereas both paths cross the edge state (in the POD space, to good approximation) and thereafter stay close to each other until reaching the AMOC-off state.

Theoretically, the MTP is expected to coincide with the instanton only in the limit of weak noise, in which case the instanton is the solution of a minimization problem involving the so-called FW action~\citep{freidlin_random_1998}.
For finite noise, deviations from the instanton are possible, particularly in regions of state space where the drift field $f$ is positively divergent~\citep{borner_saddle_2024} (i.e., $\nabla \cdot f (x)> 0$).
This effect is incorporated in the minimization of the Onsager-Machlup (OM) action, which contains a finite-noise correction to the FW action~\citep{pinski_transition_2010}.
The OM minimization problem penalises path segments located in regions of positive divergence, favoring paths that avoid these regions as much as possible~\citep{borner_saddle_2024}.

We will show here that the observed deviation between the MTP and the instanton indeed occurs in a region of positive divergence.
To estimate $\nabla \cdot f$ along the instanton, we sample points uniformly along its path in the POD latent space.
At each point, 500 initial model states are generated by linearly combining the $l$ dominant POD modes with a small perturbation:
\begin{equation}
\mathbf{X}_{0} = \sum_{k=1}^{l} (\mathbf{v}_k^{inst} + \epsilon_k) \mathbf{U}_k,
\end{equation}
where the $\mathbf{v}_k^{inst}$ are the POD coefficients of a given state belonging to the instanton and the $\epsilon_k$ are small random perturbations designed to uniformly sample an $r=0.001$ $l$-dimensional sphere around $\mathbf{v}^{inst}$.
The model is then evolved deterministically (without noise) from these 500 initial solutions for fifty time steps.
At each time step, the divergence is estimated by computing the trace of the system Jacobian, averaged over the first five time steps:
\begin{equation}
\overline{\nabla \cdot f}\ \simeq\ \mathrm{Tr}(\overline{J_{FD}}),
\end{equation}
where $\overline{J_{FD}}$ denotes the averaged Jacobian of the system, computed by finite differences (between each point and its nearest neighbour at the beginning of the time step).

\begin{figure}[]
\centering
\includegraphics[width=0.5\linewidth]{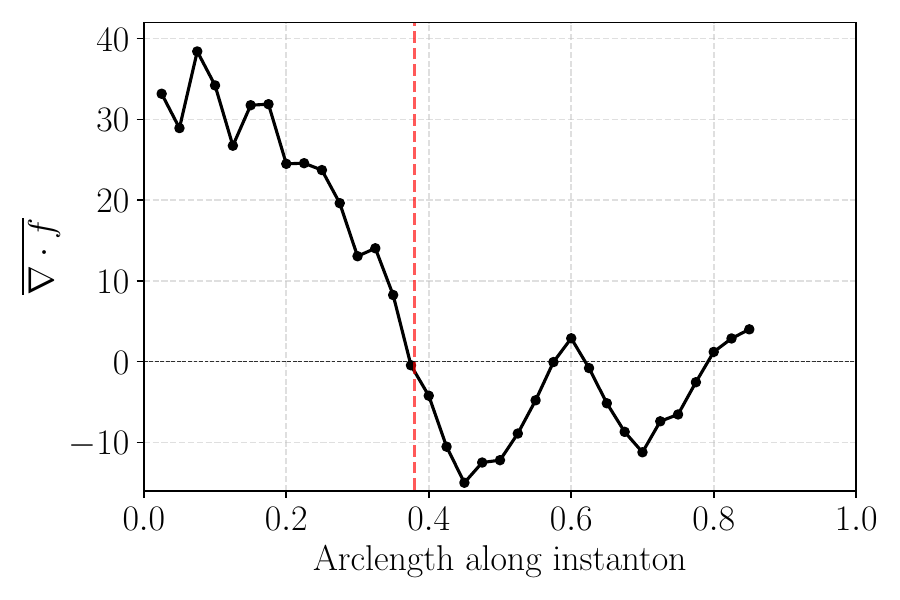}
\caption{Estimated values of $\overline{\nabla \cdot f}$ along the instanton. The red vertical dashed line indicates the position of the edge state.}
\label{fig:div_f_estimate}
\end{figure}

Along the instanton, the estimated divergence of $f$ is positive for the first third of the path (in arclength coordinates) and becomes mostly negative thereafter (Fig.~\ref{fig:div_f_estimate}).
This agrees with the theoretical argument as to where possible deviations between the MTP and instanton are expected to occur.

\bibliographystyle{aipauth4-2}
\bibliography{better_biblio.bib}

\end{document}